\documentclass[aps,prd,twocolumn,groupedaddress,showpacs,nofootinbib]{revtex4}
\usepackage{graphicx,dcolumn,bm,amssymb,amsmath,latexsym,amsfonts,footnote}
\usepackage[usenames]{color} 

\begin{document}

\newcommand{\nonu}{\nonumber}
\newcommand{\sm}{\small}
\newcommand{\noi}{\noindent}
\newcommand{\npg}{\newpage}
\newcommand{\nl}{\newline}
\newcommand{\bp}{\begin{picture}}
\newcommand{\ep}{\end{picture}}
\newcommand{\bc}{\begin{center}}
\newcommand{\ec}{\end{center}}
\newcommand{\be}{\begin{equation}}
\newcommand{\ee}{\end{equation}}
\newcommand{\beal}{\begin{align}}
\newcommand{\eeal}{\end{align}}
\newcommand{\bea}{\begin{eqnarray}}
\newcommand{\eea}{\end{eqnarray}}
\newcommand{\bnabla}{\mbox{\boldmath $\nabla$}}
\newcommand{\univec}{\textbf{a}}
\newcommand{\VectorA}{\textbf{A}}
\newcommand{\Pint}

\title{Energy extraction in electrostatic extreme binary black holes}

\author{A. Baez$^{1,}$\footnote{jbaez@fis.cinvestav.mx}, Nora Breton  $^{1,}$\footnote{nora@fis.cinvestav.mx}, and I. Cabrera-Munguia $^{2,}$\footnote{icabreramunguia@gmail.com}}
\affiliation{$^{1}$ Departamento de F\'isica, Centro de Investigaci\'on y de Estudios
 Avanzados del I. P. N.  Apdo. Postal 14-740, Mexico City, Mexico\\
$^{2}$Departamento de F\'isica y Matem\'aticas, Universidad Aut\'onoma de Ciudad Ju\'arez, 32310 Ciudad Ju\'arez, Chihuahua, M\'exico}



\begin{abstract}
Relying on the Penrose process mechanism, we study the possibility of energy extraction from a binary system composed of two extreme electrostatic black holes (BHs) oppositely charged, separated by a strut described by Bonnor's metric (BM). We determined and plotted the generalized ergosphere that surrounds only one of the BH. We demonstrate the existence of non closed orbits of negative energy outside the event horizon; these orbits allow the possibility of energy extraction by particle disintegration from a system described by the BM. Besides we prove that the extraction process can occur when a charged test particle and the BH have opposite charges;  also, we analyzed the efficiency of the process.
\end{abstract}

\maketitle

\section{Introduction}
\vspace{-0.3cm}
The Kerr metric is a stationary solution of Einstein's field equation that gives the more general description of a rotating BH \cite{Kerr1963,Visser2007,Teukolsky2015}; this metric possesses an interesting region called the ergosphere, delimited by the stationary limit surface and the outer event horizon. Within the ergosphere the timelike Killing vector $(\partial_{t})^{\mu}$  becomes spacelike, this special feature allows that particles inside the ergosphere can have negative energy. However, the particle can yet avoid enter the event horizon and can escape back to infinity.

The Penrose process is a mechanism proposed by Penrose and Floyd \cite{PenFloyd1971} for extracting energy from a rotating BH taking advantage of the fact that test particles inside the ergosphere can have negative energy states. It consists of a particle that reaches the ergosphere and at some point it disintegrates into two fragments, one of the fragments is trapped within the ergosphere with negative energy and the other one escapes back to infinity with more energy than the one of the incident particle;  by conservation of energy then rotating energy through angular momentum has been extracted. However, the ergosphere is a characteristic region of stationary solutions, then by means of the Penrose process it is, in principle, impossible to extract energy from a static BH. However, for electrostatic BHs it is possible to define a region where charged test particles can have negative energy  \cite{DenardoRuffini1972,Dadich1980,Christo1971}.

Although the Penrose process applied to a single particle might seem unfeasible to carry out, it is possible to establish relations between the Penrose process and some astrophysical observations. For instance, the collisional Penrose process might eventually eliminate dark energy particles in the vicinity of a supermassive BH once the multiple particles that scatter inside the ergosphere achieve an arbitrarily high center of mass energy \cite{Schnittman2018}. On the other hand, the influence of an external magnetic field surrounding a rotating BH can form accretion disks of charged ionized matter \cite{Kolos2021} and this may be related to high-frequency oscillations noticed in microquasars, galactic nuclei, or even the magnetic Penrose process itself \cite{Dadhich2018} where ultra high-energy particles around rotating magnetized BHs are created \cite{Kolos2017,Kolos2015}. Moreover, the radiative Penrose process is connected to synchrotron radiation of charged particles moving within the ergosphere of a magnetized BH, where such a process considers a special type of radiated photons having negative energy relative to a distant observer \cite{Stuchlik2021,Kolos2018}. In addition, recent numerical studies on plasmas and jets suggest the main role of negative energy particles and the Penrose process in the total flux coming from the BH jets \cite{Parfrey2019}. Finally, the electromagnetic Penrose process \cite{Bhat1985,Tursunov2021,Parthasarathy1986,Wagh1985,Wagh1989,Nucamendi2022} allows events of high energy emission in contrast with the efficiency of 20.7$\%$ of the well-known Penrose process due only to the rotation of a Kerr BH.

The present paper aims to investigate energy extraction via the Penrose process in a BH binary system.  In \cite{Richartz2021} the energy extraction is analyzed for the Majumdar-Papapetrou (MP) BH binary metric \cite{Majumdar,Papapetrou,Hartle1972}, which is an exact solution of the Einstein-Maxwell equations describing two static charged BHs whose charges equal their masses, $|Q_{i}|=M_{i}$, $i=1,2$. Therefore, in the MP binary metric, the BHs remain in equilibrium since their mutual gravitational attraction compensates their mutual electric repulsion no matter how far apart the sources are. Our purpose is to use the method developed in \cite{DenardoRuffini1972,Christo1971} to determine negative energy states for charged test particles and prove that by means of the Penrose process is possible the energy extraction in a binary system composed of two electrostatic oppositely charged BHs described by Bonnor's metric (BM) \cite{Bonnor1979,Cabrera2011}, where now the gravitational attraction does not counterbalance the electrical (attractive) force, and therefore, a conical singularity arises \cite{BachW,Israel} in between the sources. In particular we study the generalized ergosphere, its dependence on charged test particle and how the energy extraction efficiency is affected by the presence of a BH companion with opposite charge.

Our paper is organized as follows. In Sec. \ref{sec:geodesics}, we introduce the spacetime described by the BM and derive the motion equations for charged massive test particles in a static and axisymmetric spacetime. In Sec. \ref{sec:ergos} the generalized ergosphere and the existence of negative orbits is analyzed. In Sec. \ref{sec:penrose} the Penrose process is described and the conservation equations are presented (mass, charge, energy, linear momentum and angular momentum). In Sec. \ref{sec:eficiencia} the constrictions over the parameters and the maximum efficiency of the process are presented and in the last section conclusions are given.

\section{Bonnor's Binary BH}
\vspace{-0.3cm}
The stationary axisymmetric line element in Weyl's cylindrical coordinates $(t, \rho, z, \phi)$ is the Papapetrou metric \cite{Papapetrou} given by
\be ds^{2}=f^{-1}\big[e^{2\gamma}(d\rho^{2}+dz^{2})+\rho^{2}d\phi^{2}\big]- f(dt-\omega d\phi)^{2},
\label{papa}\ee
\noi where the condition $\omega=0$ defines the double Reissner-Nordstr\"{o}m  (DRN) \cite{Manko2007} solution of Einstein-Maxwell equations that describes a two-body system composed of two electrostatic BHs; to keep the two BHs apart, a line source should be introduced, the  {\it strut} \cite{BachW,Israel}. As a consequence there arise an angle deficit in $\phi$ that is a conical singularity. The interacting force associated to the strut between the BHs is given by \cite{Manko2007}
\begin{align} \mathcal{F}&=\frac{M_{1}M_{2}-(Q_{1}-\delta)(Q_{2}+\delta)}{R^{2}-(M_{1}+M_{2})^{2}+(Q_{1}+Q_{2})^{2}},\nonu\\
\delta&=\frac{M_{2}Q_{1}-M_{1}Q_{2}}{R+M_{1}+M_{2}}, \label{Manko} \end{align}
\noi where $M_{i}$ and $Q_{i}$, for $i=1,2,$ are the masses and electric charges, respectively, while $R$ defines an arbitrary separation distance among the BH centers. Regarding the last point, the half-length BH horizons $\sigma_{i}$ assume the next simple formulas \cite{VCH2002,Alekseev2007}
\vspace{-0.1cm}
\begin{align} \sigma_{i}&=\sqrt{M_{i}^{2}- Q_{i}^{2}-2(-1)^{i}Q_{i}\frac{M_{2}Q_{1}-M_{1}Q_{2}}{R+M_{1}+M_{2}}}, \nonu\\ i&=1,2, \label{VarzChist} \end{align}
\noi and thus the DRN spacetime contains a total, of five independent parameters within the set $\{M_{i},Q_{i},R\}$. The reader should note that if one exchanges the physical properties of the sources; i.e.,  $M_{1} \leftrightarrow M_{2}$ and $Q_{1} \leftrightarrow Q_{2}$, where $\delta \to-\delta$, the interaction force remains invariant. Also, it is possible to observe that $\sigma_{1} \leftrightarrow \sigma_{2}$ under this physical interchange.

The extremal condition $\sigma_{i}=0$ leads to two possible scenarios. The first one is the MP case, where $|Q_{i}|=M_{i}$, and there is no need of introducing the strut since
the gravitational attraction balances the electric repulsion and the net force is null, $\mathcal{F}=0$.

The second scenario corresponds to the Bonnor's metric (BM); in this case the electric charges, $Q_{i}$, are opposite in sign and related to the masses and distance according to \cite{Cabrera2011}
\begin{equation}
\begin{aligned}
Q_{1} = M_{1}\sqrt{\frac{(R+M_{2})^{2}-M_{1}^2}{(R-M_{2})^{2}-M_{1}^2}}, \\
Q_{2} =- M_{2}\sqrt{\frac{(R+M_{1})^{2}-M_{2}^2}{(R-M_{1})^{2}-M_{2}^2}},
\label{chargesB} \end{aligned}
\end{equation}
\noi where the electric charges and masses fulfill the condition $|Q_{i}|>M_{i}$, $i=1,2$. The metric functions in the BM are given by
\begin{widetext}
\begin{align}
f &= \bigg( \frac{1}{1+g_{1}}+\frac{1}{1+g_{2}}-1\bigg)^{2}, \qquad e^{2\gamma} =\bigg(\frac{(1-g_{1} g_{2}) g_{+}g_{-}}{4d^2}\bigg)^{4}, \nonu\\
g_{1}&=\frac{(1+d)M_{1}-(1-d)M_{2}}{(1+d)r_{-}-(1-d)r_{+}},\quad g_{2} =  \frac{(1+d)M_{2}-(1-d)M_{1}}{(1+d)r_{+}-(1-d)r_{-}}, \qquad d =\sqrt{\frac{R^{2}-(M_{1}-M_{2})^{2}}{R^{2}-(M_{1}+M_{2})^{2}}} \nonu\\
g_{+}&=1+d-(1-d)\frac{r_{+}}{r_{-}}, \qquad
g_{-}=1+d-(1-d)\frac{r_{-}}{r_{+}},  \qquad
r_{\pm} = \sqrt{\rho^{2}+(z\pm R/2)^2}.
\label{bmmetric}
\end{align}
\end{widetext}

In Weyl's cylindrical coordinates the BHs and their horizons are represented by points at the $z$-axis $(\rho=0, z=\pm R/2)$. The independent parameters of the metric functions in the BM (\ref{bmmetric}) are $M_{1}$, $M_{2}$, and $R$, while the electric potential $A_{\mu}$ is given by
\begin{equation}\label{potential}
A_{\mu} = (A_{t}, A_{\rho}, A_{\phi},A_{z}) = \left(\frac{1}{1+g_{1}}-\frac{1}{1+g_{2}},0,0,0\right).
\end{equation}

Contrary to the MP scenario, in the BM the electrical force is attractive rather than repulsive, which means that the gravitational attraction will never balance the electromagnetic force. After replacing (\ref{chargesB}) in (\ref{Manko}) it is possible to show that the interaction force associated to the strut between the BHs is given by \cite{Cabrera2014}
\begin{equation}
\mathcal{F}=\frac{2 M_{1}M_{2}}{R^{2}-\left(M_{1}+M_{2}\right)^{2}}\left(1+\frac{2M_{1}M_{2}}{R^{2}-\left(M_{1}+M_{2}\right)^2}\right),
\end{equation}

\noi and because the masses and distance satisfy the inequality $(M_{1}+M_{2})<R$, only attractive scenarios will be allowed corresponding to a positive force of the strut, $\mathcal{F}>0$. On the other hand, since the BM is static, the spacetime does not possess an ergosphere in the usual sense where the timelike Killing vector becomes spacelike, and  consequently the energy associated to geodesic motion of neutral particles is always positive, i.e., the energy extraction is not possible. However, charged particles can interact with charged BHs via Lorentz forces. According to the ideas proposed by Denardo-Ruffini \cite{DenardoRuffini1972} and Dadhich \cite{Dadich1980} for a single charged BH as well as Sanches-Richartz for a BH binary \cite{Richartz2021}, we can define a particular region where negative energy trajectories and energy extraction are, in principle, possible. In what follows we study the motion of charged particles in the BH binary BM  and derive the energy extraction.

\vspace{-0.2cm}
\subsection{Motion of charged particles}\label{sec:geodesics}
\vspace{-0.2cm}
The motion equations for a test particle with charge-mass ratio $\mu$ in a spacetime characterized by the metric $g_{\mu \nu}$ and interacting with the electric potential $A_{\mu}$, can be obtained from the Euler-Lagrange equations with the Lagrangian
\begin{equation}
\mathcal{L} = \frac{1}{2} g_{\mu\nu}\dot{x}^{\mu}\dot{x}^{\nu} + \mu A_{\alpha}\dot{x}^{\alpha},
\end{equation}

\noi where the dot means derivative with respect to an affine parameter. In terms of the metric coefficients in Eq. (\ref{papa}), the Lagrangian is
\begin{equation}\label{lagrangian}
\mathcal{L} = \frac{1}{2}\left(f^{-1}\big(e^{2\gamma}(\dot{\rho}^{2}+\dot{z}^{2})+\rho^{2}\dot{\phi}^{2}\big)-f \dot{t}^{2}\right) + \mu A_{t} \dot{t},
\end{equation}

\vspace{-0.1cm}
\noi which does not depend explicitly on the coordinates $t$ and $\phi$. Then we can identify two motion constants of the test particle: its energy $E$ and angular momentum $L$ per unit mass, as measured by observers at infinity, given by
\begin{equation} \label{constants}
\begin{aligned}
E &= - \frac{\partial \mathcal{L}}{\partial \dot{t}} =  f \dot{t}-\mu A_{t},\\
L &= \frac{\partial \mathcal{L}}{\partial \dot{\phi}} = f^{-1}\rho^{2} \dot{\phi},
\end{aligned}
\end{equation}

\vspace{-0.1cm}
\noi where after solving for $\dot{\phi}$ and $\dot{t}$,  and substituting the result into Eq. (\ref{lagrangian}), we obtain the motion equations for $\rho$ and $z$ coordinates, via the Euler-Lagrange equations, as
\begin{widetext}
\begin{equation}\label{rhozeta}
\begin{aligned}
&\ddot{\rho} - \frac{f e^{-2\gamma}}{2}\frac{\partial}{\partial\rho}\left(\frac{(E+\mu A_{t})^2}{f}-\frac{f L^{2}}{\rho^{2}}\right) +\left( \frac{\dot{z}^{2}-\dot{\rho}^{2}}{2}\right)\frac{\partial}{\partial\rho}\Big(\ln\left(f e^{-2\gamma}\right)\Big) - \dot{\rho} \dot{z} \frac{\partial}{\partial z}\bigg(\ln\left(f e^{-2\gamma}\right)\bigg)=0,\\
&\ddot{z} - \frac{f e^{-2\gamma}}{2}\frac{\partial}{\partial z}\left(\frac{(E+\mu A_{t})^2}{f}-\frac{f L^{2}}{\rho^{2}}\right) + \left(\frac{\dot{\rho}^{2}-\dot{z}^{2}}{2}\right)\frac{\partial}{\partial z}\bigg(\ln\left(f e^{-2\gamma}\right)\bigg)  - \dot{\rho} \dot{z} \frac{\partial}{\partial \rho}\bigg(\ln\left(f e^{-2\gamma}\right)\bigg)=0.
\end{aligned}
\end{equation}
\end{widetext}

It is worth mentioning that the motion equations for $\rho$ and $z$, in terms of the metric (\ref{papa}) are given in \cite{Dubeibe2016} for neutral test particles. On the other hand, an explicit expression for the energy $E$ is obtained by plugging the constants of motion (\ref{constants}) into the normalization condition of the four velocity $\dot{x}^{\mu}\dot{x}_{\mu}=-1$,
\vspace{-0.1cm}
\begin{equation}\label{energy}
E = -\mu A_{t} \pm \sqrt{e^{2\gamma}\left(\dot{\rho}^{2}+\dot{z}^{2}\right)+\frac{L^{2} f^{2}}{\rho^{2}}+f},
\end{equation}

\vspace{-0.2cm}
\noi where the positive root is taken for a positive energy at infinity when $\mu=0$ \cite{Christo1971}. After some straightforward algebra, Eq. (\ref{energy}) can be  expressed as
\begin{equation}\label{eqrhoz}
\dot{\rho}^{2}+\dot{z}^{2} = E_{\rm{eff}}^{2}(\rho,z)-V_{\rm{eff}}(\rho,z),
\end{equation}

\noi where
\begin{align}
E_{\rm{eff}}^2(\rho,z) &=\frac{\left(E + \mu A_{t}\right)^2}{e^{2\gamma}}, \nonu \\  V_{\rm{eff}}(\rho,z)&=\frac{1}{e^{2\gamma}}\left(\frac{L^{2}f^{2}}{\rho^{2}}+f\right).
\label{effveff}
\end{align}

These expressions are subject to the constraints:
\begin{equation}
E_{\rm{eff}}(\rho,z)\geq0, \qquad E_{\rm{eff}}^{2}(\rho,z)\geq V_{\rm{eff}}(\rho, z).
\end{equation}

\vspace{-0.1cm}
The system of equations (\ref{rhozeta}) fully describes the motion of a charged test particle in a static and axisymmetric spacetime; this set of equations can be numerically solved once the appropriate initial conditions are chosen. In order to solve this system of equations, the values for the energy $E$, angular momentum $L$, and initial values for $\rho$, $z$, and $\dot{z}$ should be given; the initial value for $\dot{\rho}$ is determined from  Eq. (\ref{energy}). With these initial values  Eqs. (\ref{rhozeta}) can be solved for  $\rho(\lambda)$ and $z(\lambda)$ with $\lambda$ being the affine parameter; the full description of the motion of a test particle is obtained when the set of Eqs. ($\ref{constants}$) is solved using  $\rho(\lambda)$ and $z(\lambda)$.

\vspace{-0.2cm}
\subsection{Generalized ergosphere}\label{sec:ergos}
\vspace{-0.2cm}
 From the expression (\ref{energy}), we know that the energy is determined by the angular momentum $L$, charge-mass ratio $\mu$, electric potential $A_{t}$ and the coordinates and velocities: $\rho$, $z$, $\dot{\rho}$ and $\dot{z}$ at a specific time. In order to know if there are test particles with negative energy we consider the minimum possible energy with a fixed charge $\mu$ and position $(\rho ,z)$;  this is the energy associated with test particles at rest. Replacing $\dot{\rho}=0$, $\dot{z}=0$, and $L=0$ in (\ref{energy}), we get
 \begin{equation}\label{emin1}
 E_{\rm{min}} = - \mu A_{t} + \sqrt{f},
 \end{equation}

\noi since $\sqrt{f}\geq0$, then the existence of test particles with negative energies is defined by the term $- \mu A_{t}$. In order that $E_{\rm{min}}<0$, $\mu A_{t}>0$. Hence we need to determine the sign of $A_{t}$ given by,
\begin{equation}
A_{t}= \frac{1}{1+g_{1}}-\frac{1}{1+g_{2}}.
\end{equation}

Using the explicit form of $g_{1}$ and $g_{2}$ in Eq. (\ref{bmmetric}) one can easily verify that the sign of $A_{t}$ is different if $(\rho, z)$ are inside or outside the circle delimited by
\begin{equation}\label{circle}
\bar{\rho}^{2}+\bigg(\bar{z}+\frac{1}{2}\left(\frac{1+M_{R}^{2}}{1-M_{R}^{2}}\right)\bigg)^{2} =\left(\frac{M_{R}}{1-M_{R}^2}\right)^2,
\end{equation}

\noi where $\bar{\rho}=\rho/R$, $\bar{z}=z/R$ and $M_{R}=M_{2}/M_{1}$. Note that this region encircles the smaller mass. We distinguish two scenarios: \nl
i) For $M_{1}>M_{2}$ the circle represented by (\ref{circle}) surrounds the BH with mass $M_{2}$ and electric charge $Q_{2}<0$. Whether $(\rho, z)$ is located inside (outside) this circle, then, it follows that $A_{t}>0 (A_{t}<0)$.
\noi ii) For $M_{1}<M_{2}$, the circle depicted by (\ref{circle}) surrounds the BH with mass $M_{1}$ and electric charge $Q_{1}>0$. If $(\rho, z)$ is localized inside (outside) the circular region, then $A_{t}>0 (A_{t}<0)$.

It should be stressed that the region (\ref{circle}) never includes both sources. For the case $M_{1}>M_{2}$, the circle remains on the semi-plane $z<0$ while for $M_{2}>M_{1}$ it is located on the semi-plane $z>0$. Hence, for a given value $\mu>0$,  the condition $\mu A_{t}>0$ cannot be simultaneously fulfilled; in this case the ergosphere encircles the lower BH with electric charge $Q_{2}<0$. 
While if $\mu<0$, the condition $\mu A_{t}>0$ is fulfilled in the region encircling the upper BH with electric charge $Q_{1}>0$.

\begin{figure*}[ht]
\includegraphics[width=6.0cm,height=6.0cm]{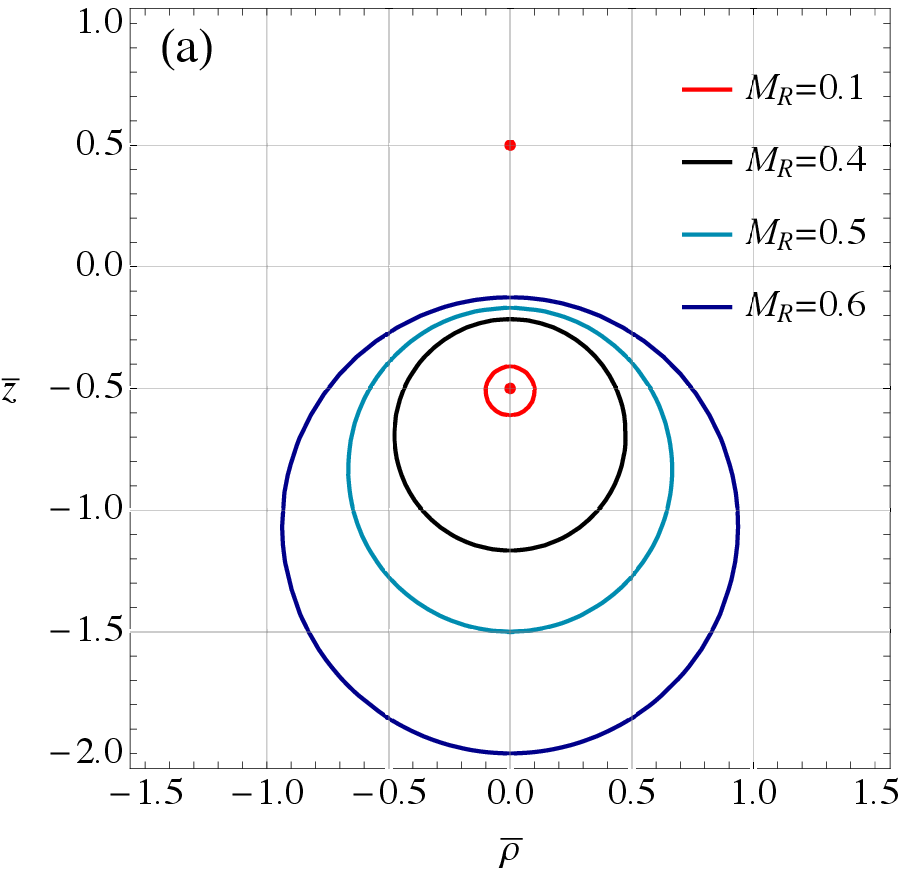}
\quad
\includegraphics[width=6.0cm,height=6.0cm]{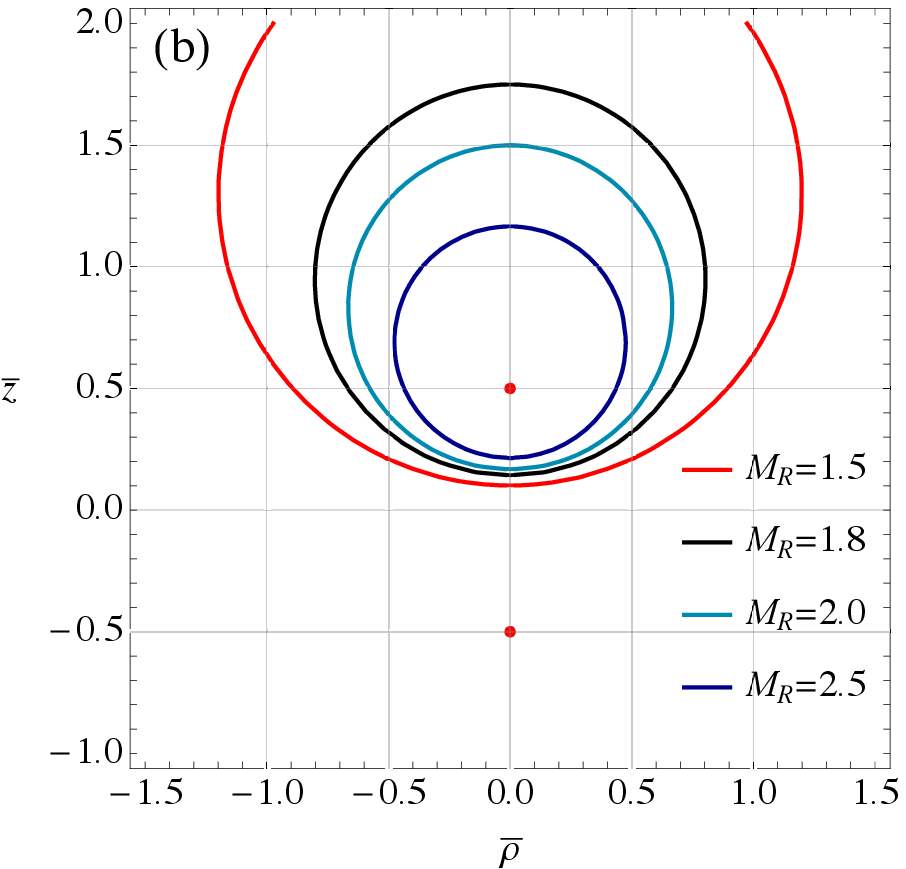}
\caption{\label{fig:circle1} The sign of the electric potential, $A_{t}$ depends on the chosen $(\rho, z)$; (a) If $M_{1}>M_{2}$, $0<M_{R}<1$ then $A_{t}$ is positive inside the circle that surrounds the bottom source, and negative outside. (b) On the other hand, if $M_{1}<M_{2}$, $M_{R}>1$ then $A_{t}$ is negative inside the circle that surrounds the upper source, and positive outside.}
\end{figure*}

\begin{figure*}[ht]
\includegraphics[width=4.0cm,height=4.0cm]{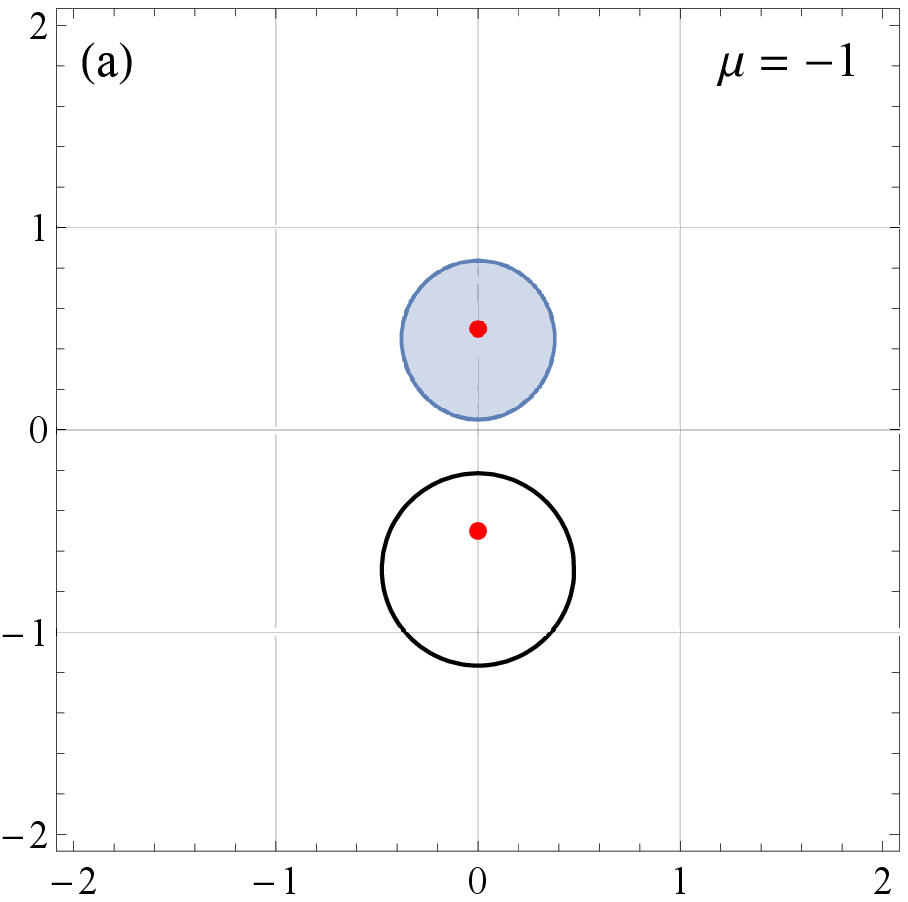}
\quad
\includegraphics[width=4.0cm,height=4.0cm]{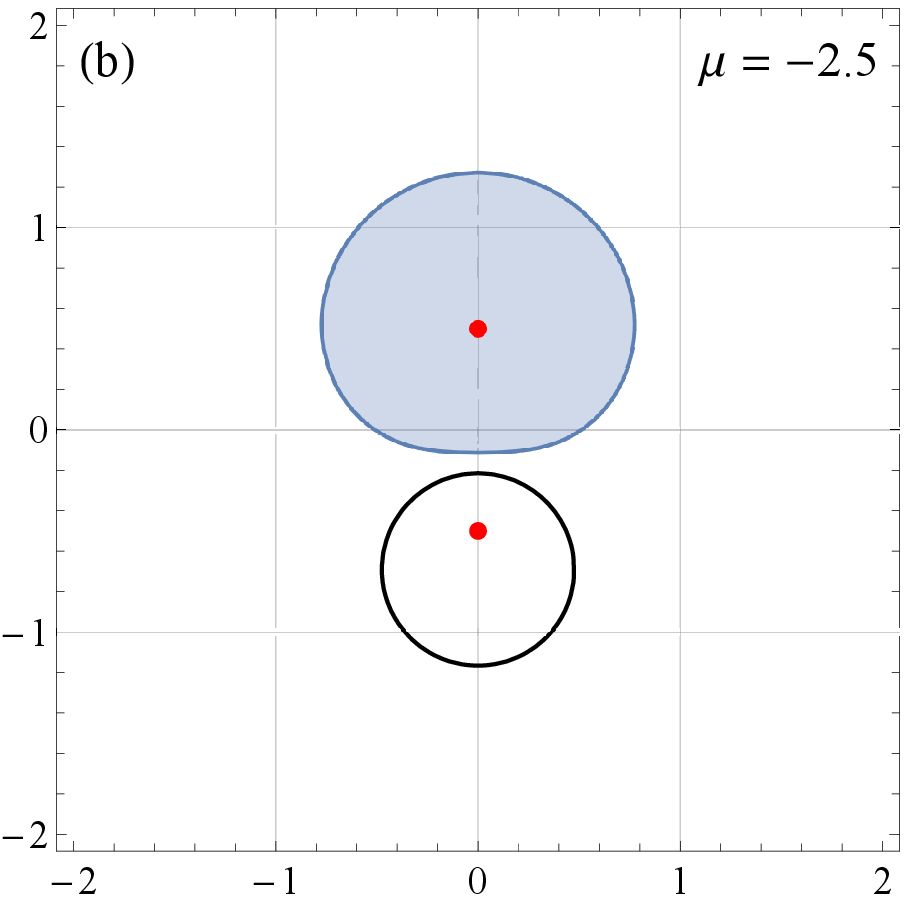}
\quad
\includegraphics[width=4.0cm,height=4.0cm]{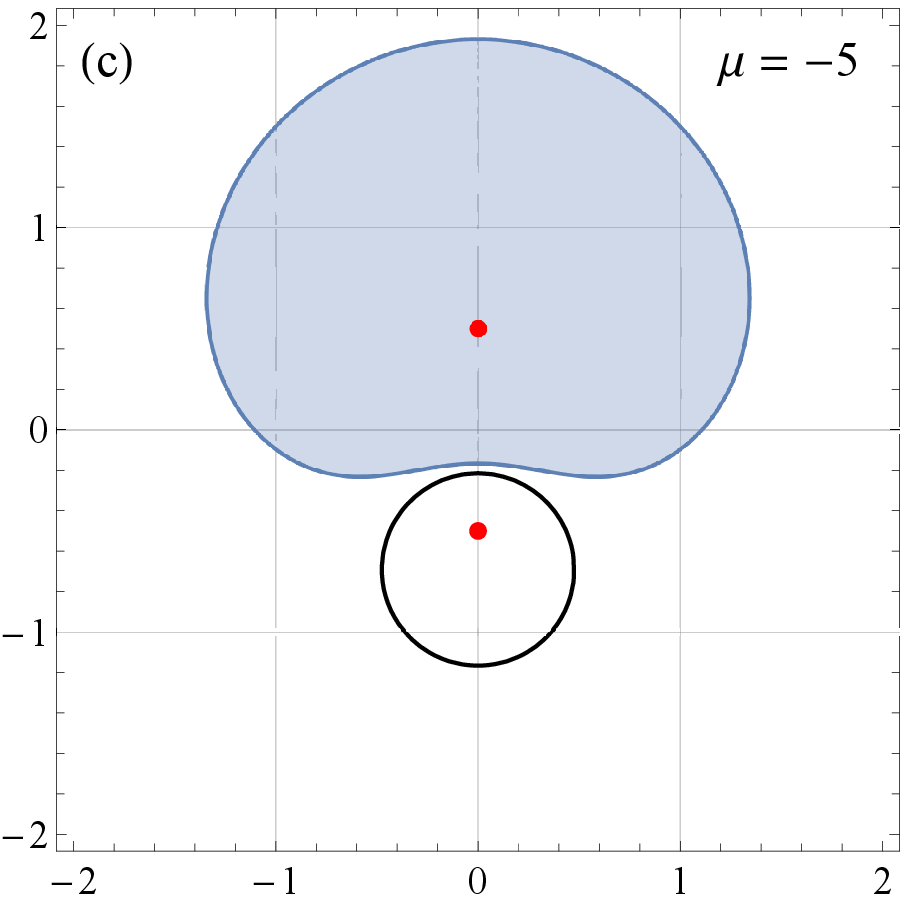}
\quad
\includegraphics[width=4.0cm,height=4.0cm]{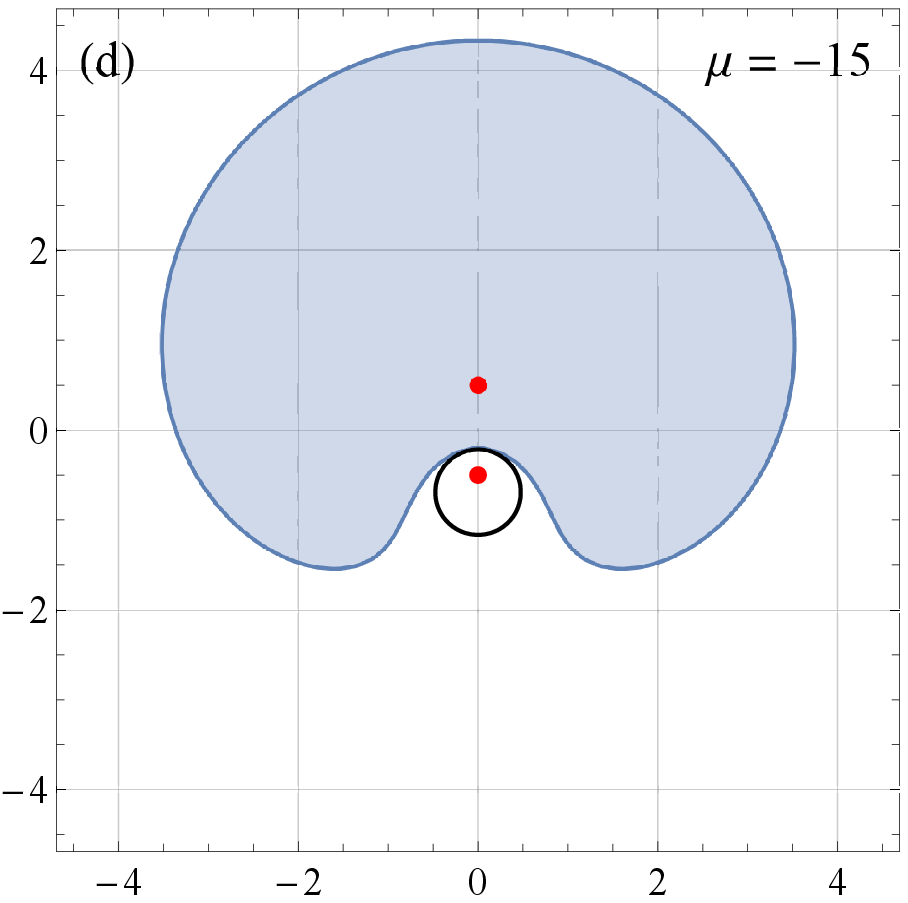}
\quad
\includegraphics[width=4.0cm,height=4.0cm]{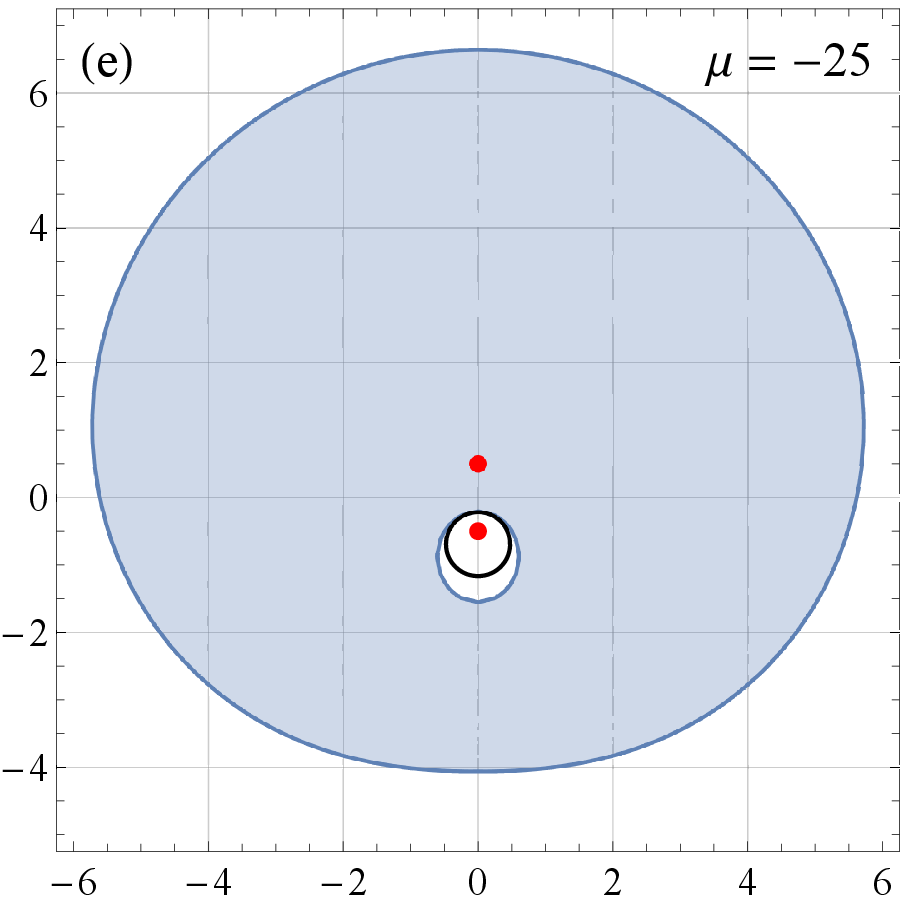}
\quad
\includegraphics[width=4.0cm,height=4.0cm]{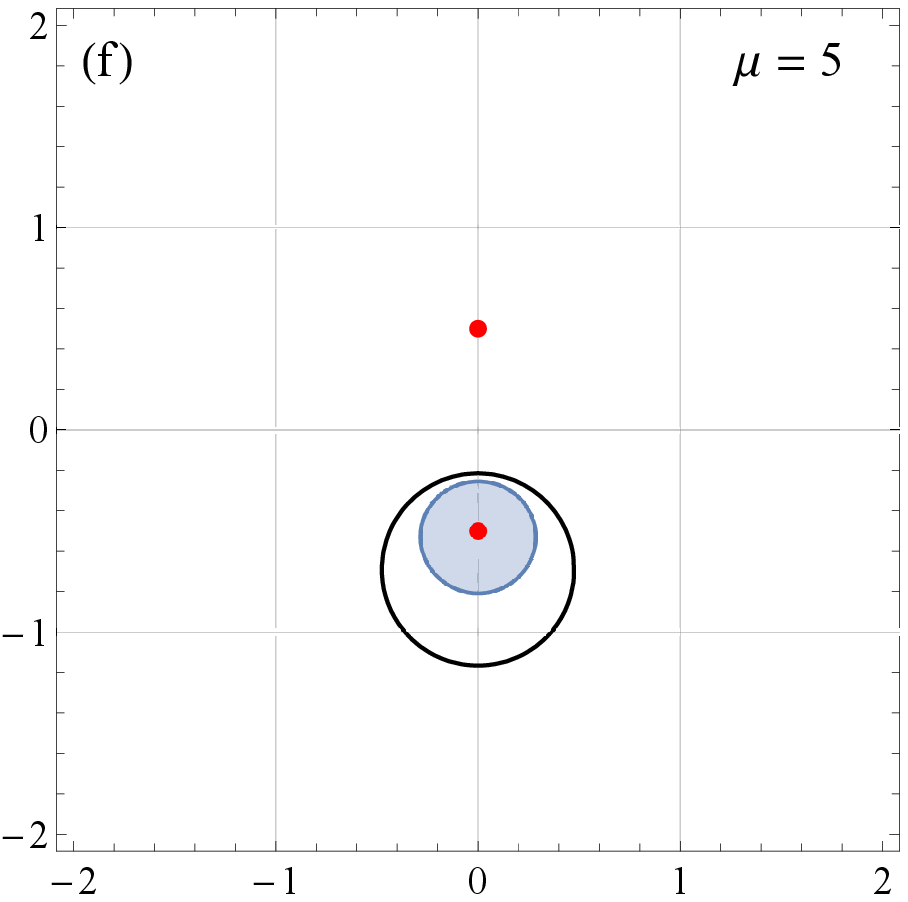}
\quad
\includegraphics[width=4.0cm,height=4.0cm]{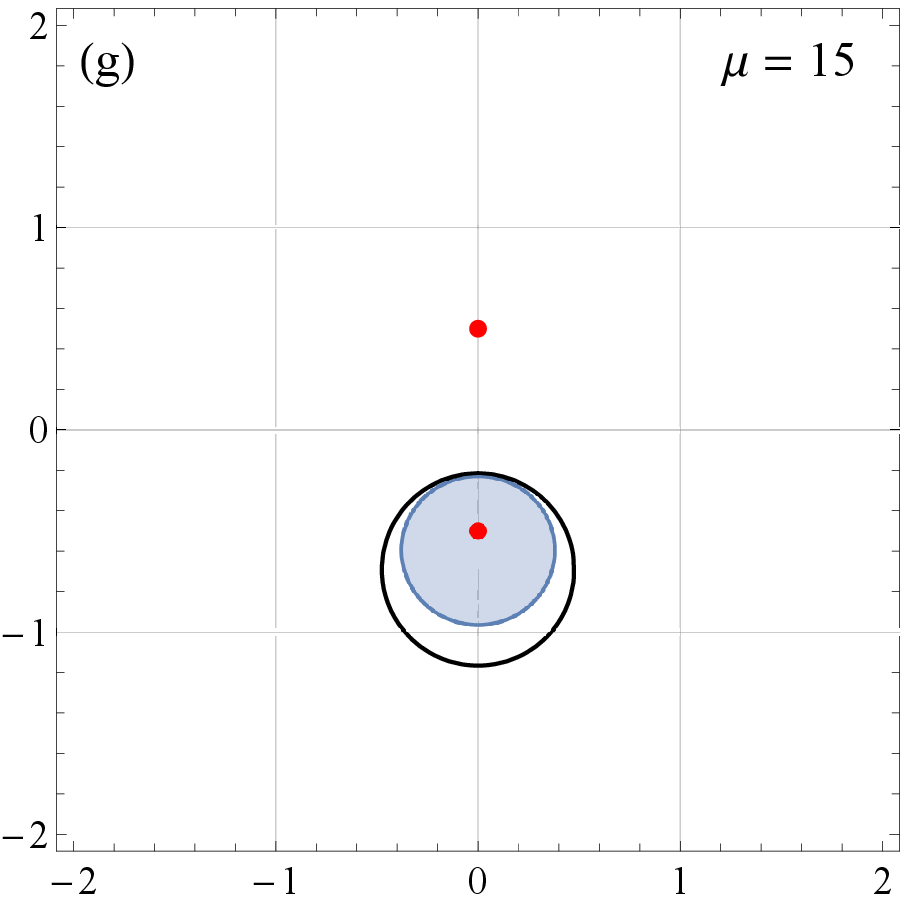}
\quad
\includegraphics[width=4.0cm,height=4.0cm]{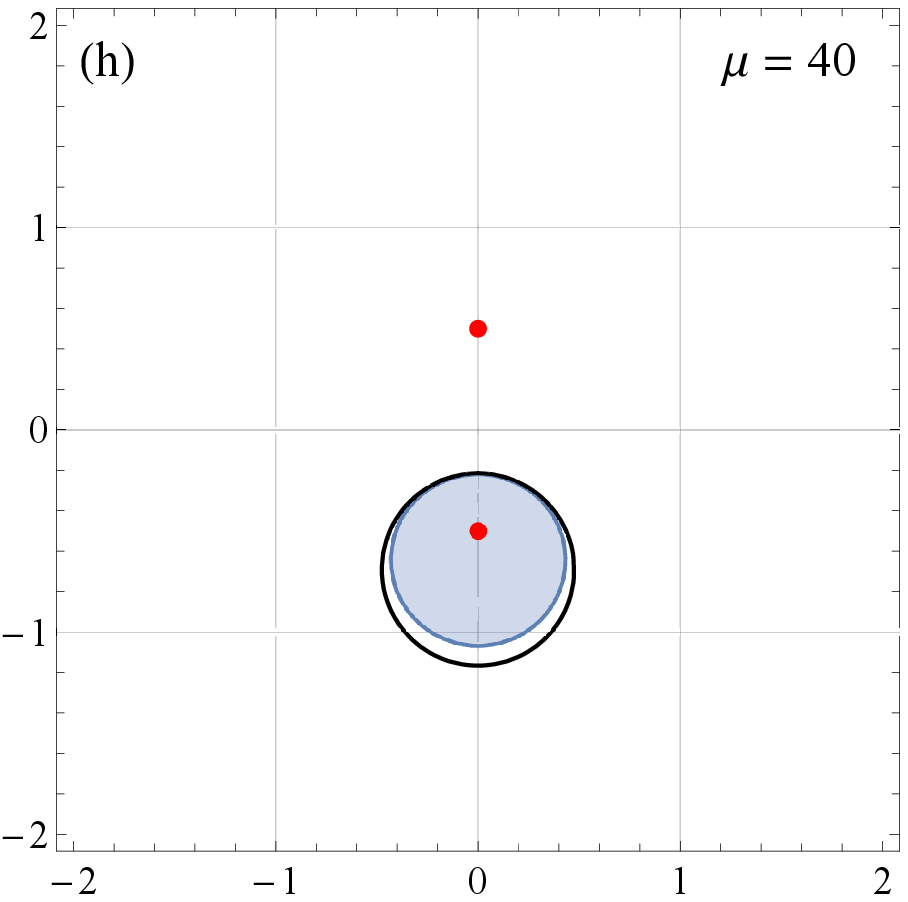}
\caption{\label{fig:ergosphere} It is shown the $\phi = 0$ (meridional) plane section of the generalized ergosphere (shaded surface) for different values of $\mu$ and fixed values of $M_{1}=0.5$, $M_{2}=0.2$, and $R=1$. Recalling that
negative energy states are achieved via the condition $\mu A_{t}>0$, where $A_{t}$ depends on $M_{1}$, $M_{2}$, and $R$. The black circumference represents Eq.\ (\ref{circle}) and the dots symbolize the BHs. The shadow regions in (a)-(e) represent the ergosphere for $\mu<0$, where the condition $\mu A_{t}>0$ is satisfied outside the black circumference, thus, the ergosphere surrounds only the upper positively charged source and excludes the region delimited by the black circumference. On the other hand, the shadow regions in (f)-(h) represent the ergosphere for $\mu>0$ that is surrounding the bottom (negatively charged) source, where now $\mu A_{t}>0$ is fulfilled inside the black circumference. For an arbitrarily large value of $|\mu|$ the ergosphere tends to occupy the space inside the black circumference. The reader must be aware that the ergosphere always surrounds the BH carrying on an opposite charge to the test particle.}
\end{figure*}

\begin{figure*}[ht]
\includegraphics[width=4.2cm,height=5.2cm]{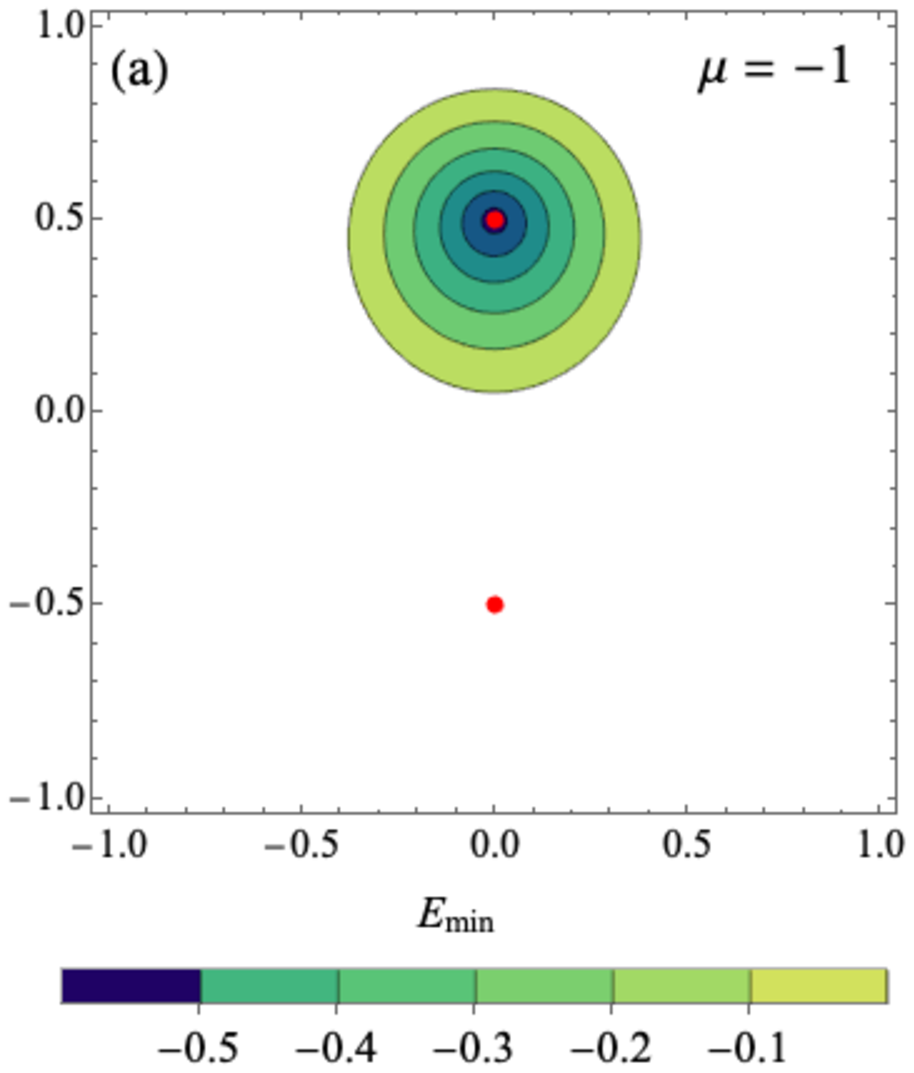}
\,
\includegraphics[width=4.2cm,height=5.2cm]{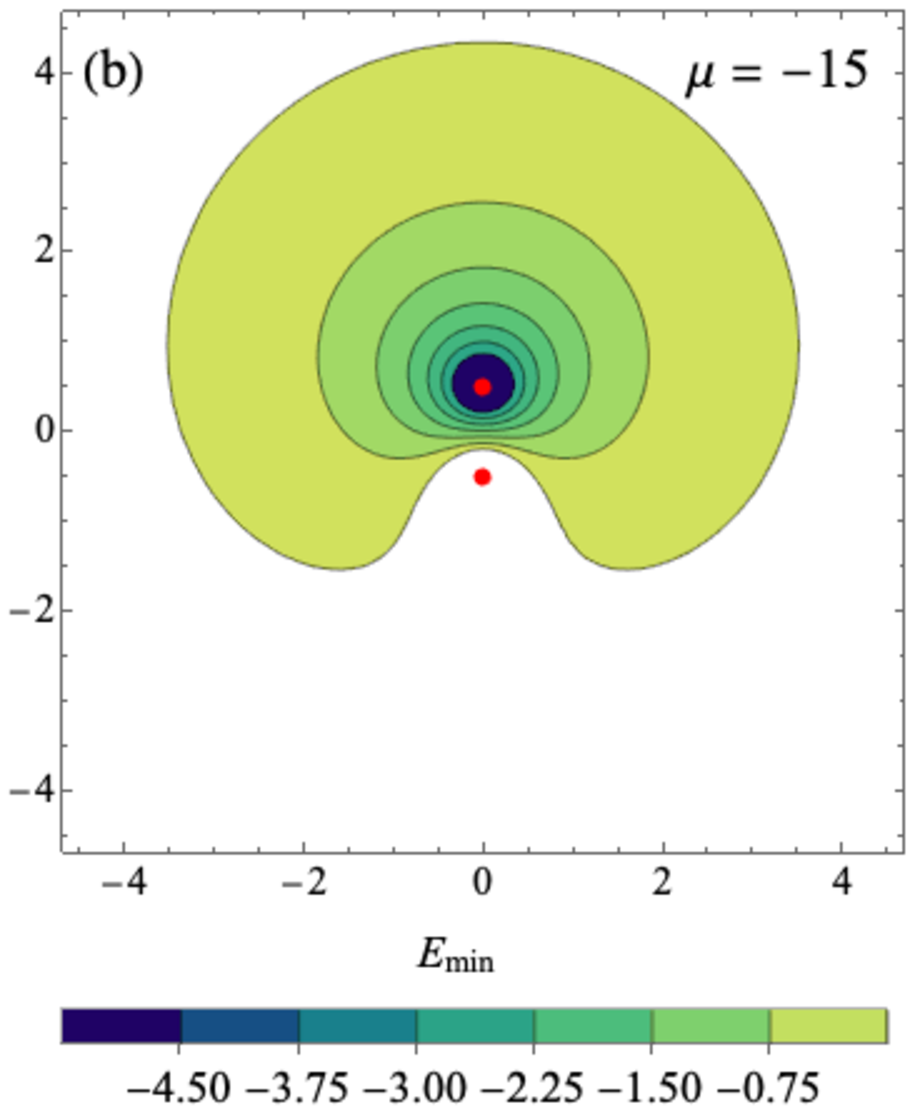}
\,
\includegraphics[width=4.2cm,height=5.2cm]{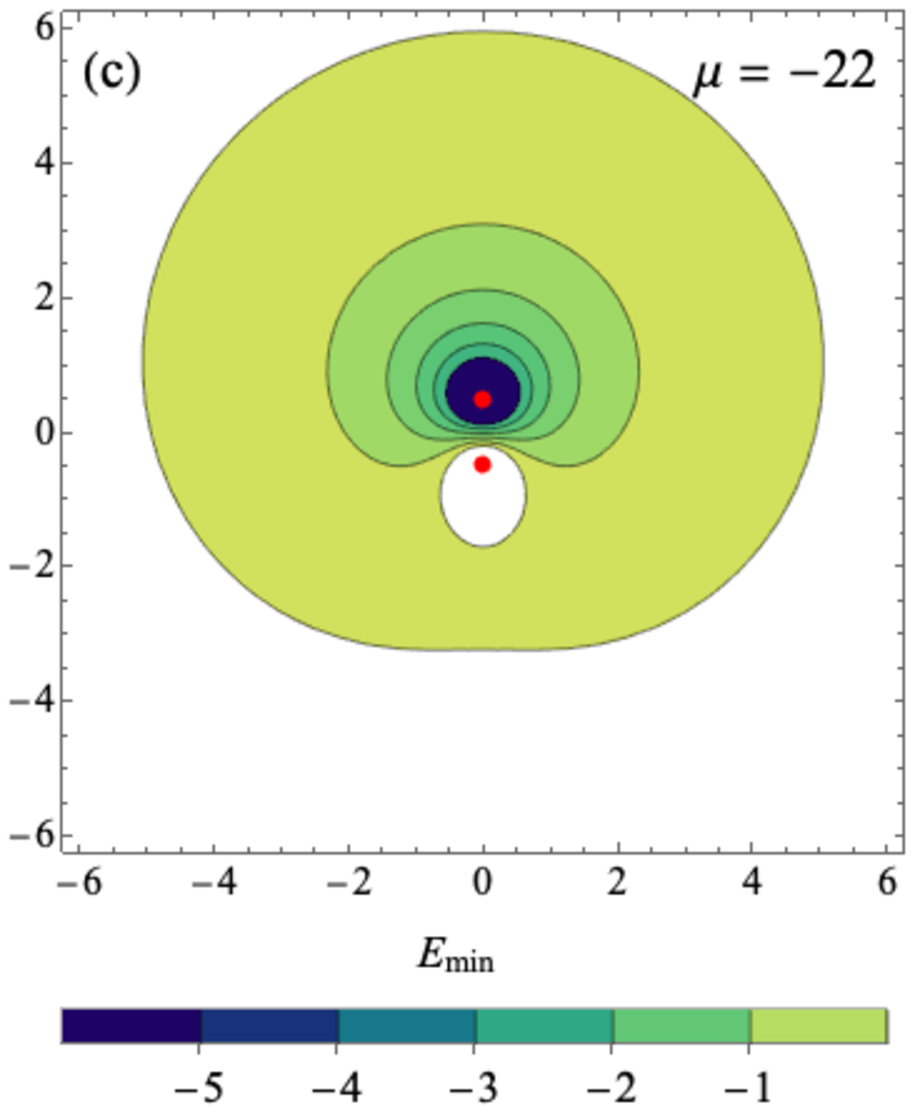}
\,
\includegraphics[width=4.2cm,height=5.2cm]{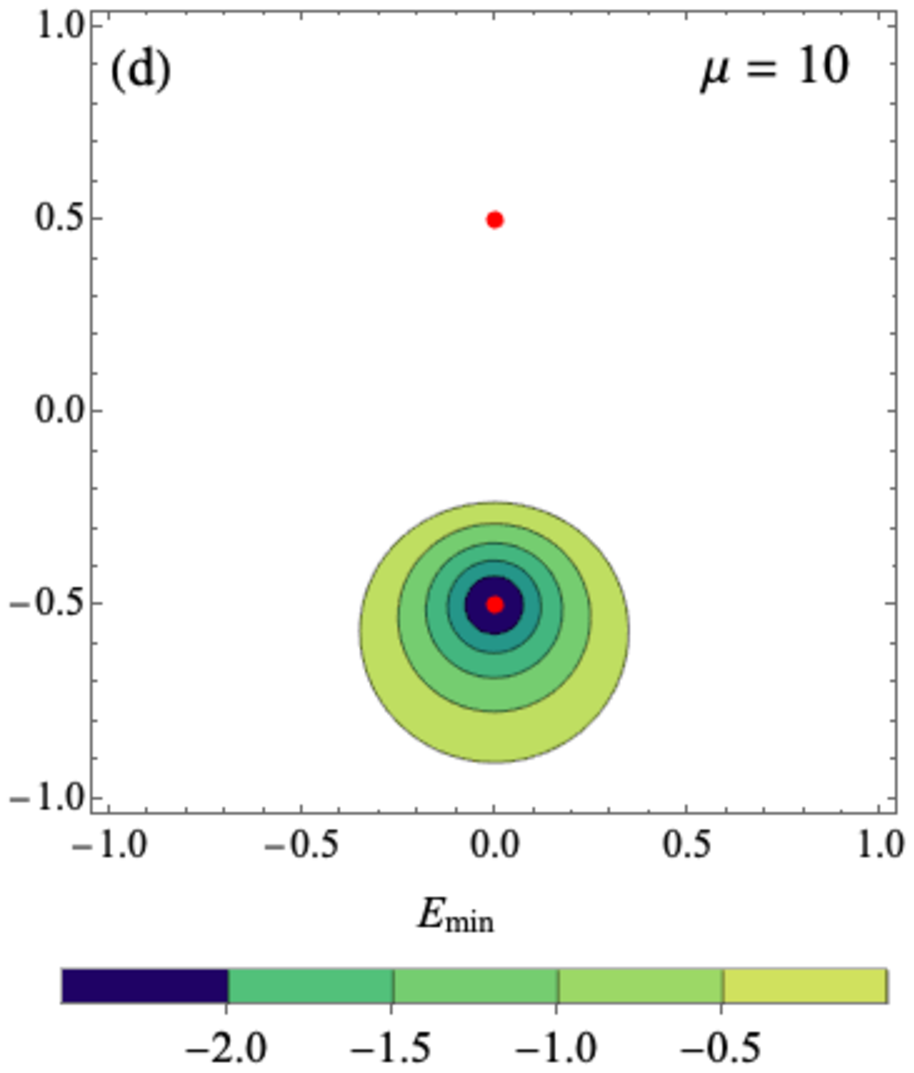}
\caption{\label{fig:level} Energy levels of the BM generalized ergosphere  are illustrated with $R=1$, $M_{1}=0.5$, $M_{2}=0.2$ and selected values of $\mu$. The color bar represents $E_{min}$. The dots indicate the location of the BHs. The horizontal and vertical axes are $\rho$ and $z$, respectively.}
\end{figure*}

When the values of $M_{1}$ and $M_{2}$ are exchanged the graphic in Fig. \ref{fig:circle1}, is rotated about the $z =0$ plane; then, without loss of generality, we shall consider only the case $M_{1}>M_{2}$.

In order for $E_{min}<0$ the sign of $\mu$ should be the same than the sign of $A_{t}$. In contrast with the MP metric, analyzed in \cite{Richartz2021}, where the two BH are positively charged and then energy can be extracted only by negatively charged test particles, in the BM
energy can be extracted by negative or positive charged test particles.

 The condition $E_{min}<0$, Eq. (\ref{emin1}) can be written as
\begin{equation}\label{ergosphere}
\bar{\rho}^2 +\bar{z}^2+\frac{2 \mu  \left(\bar{M}_{1} \bar{r}_{+}-\bar{M}_{2} \bar{r}_{-}\right) \bar{d}}{\bar{d}^2-1}+\frac{\bar{d}^{2}+1}{\bar{d}^{2}-1}\bar{r}_{+}\bar{r}_{-}<\frac{1}{4},
\end{equation}

\noi where  $\bar{r}_{\pm} =r_{\pm}/R$, $\bar{M}_{1,2}=M_{1,2}/R$ and $\bar{d}=\sqrt{\left(1-(\bar{M}_{1}-\bar{M}_{2})^{2}\right)/\left(1-(\bar{M}_{1}+\bar{M}_{2})^{2}\right)}$ are dimensionless quantities restricted by $\left(\bar{M}_{1}+\bar{M}_{2}\right)<1$. Recall that $(\bar{\rho}, \bar{z})$ are the coordinates of the initial particle, then (\ref{ergosphere}) restricts the position of the initially at rest particle. Note that this reparametrization is equivalent to fix $R=1$. The inequality (\ref{ergosphere}) determines the generalized ergosphere on the BM. The sketches of the ergosphere for $\mu>0$ and $\mu<0$ are shown in Fig.\ \ref{fig:ergosphere}; the charged test particle can have negative energy in the region {encircling the} BH with opposite charge. For particles with $\mu<0$ the generalized ergosphere surrounds the upper  source Fig. \ref{fig:ergosphere}.a-\ref{fig:ergosphere}.e. If $|\mu|$ is sufficiently large the generalized ergosphere would surround the region delimited by (\ref{circle}) without including it. On the other hand, for particles $\mu>0$ the generalized ergosphere surrounds the bottom source, Fig. \ref{fig:ergosphere}.f-\ref{fig:ergosphere}.h; and if $|\mu|$ is sufficiently large the generalized ergosphere is delimited by (\ref{circle}).

From Fig.\ \ref{fig:ergosphere} we see that the generalized ergosphere depends on the parameters $\mu$, $M_1$ and $M_{2}$, i.e., not only on the spacetime geometry but also on the charge of the test particle via $\mu$. The minimum energy per unit mass at a given point inside the ergosphere also depends on  $\mu$ and  one the masses of the BHs. To illustrate this, the energy levels of the ergosphere are shown in Fig.\ \ref{fig:level} for different values of $M_1$, $M_{2}$ for charge-mass ratio $\mu$ positive and negative. One can see that in general the shape of the ergosphere does not change and the magnitude of the energy levels is larger as the test particle gets closer to the BH oppositely charged. The ergospheres then surround one of the BHs, the one with charge opposite to the test particle and then never merge to include both BHs, in this sense resembling the double--Kerr for counterrotating BHs \cite{Herdeiro2009}.

\vspace{-0.2cm}
\subsection{Negative energy trajectories}
\vspace{-0.2cm}
The negative energy trajectories of the charged test particles in the BM are confined inside the ergosphere defined by (\ref{ergosphere}); in the case illustrated the particle falls  into one of the BHs.

\begin{figure}[ht]
\includegraphics[width=4cm,height=4cm]{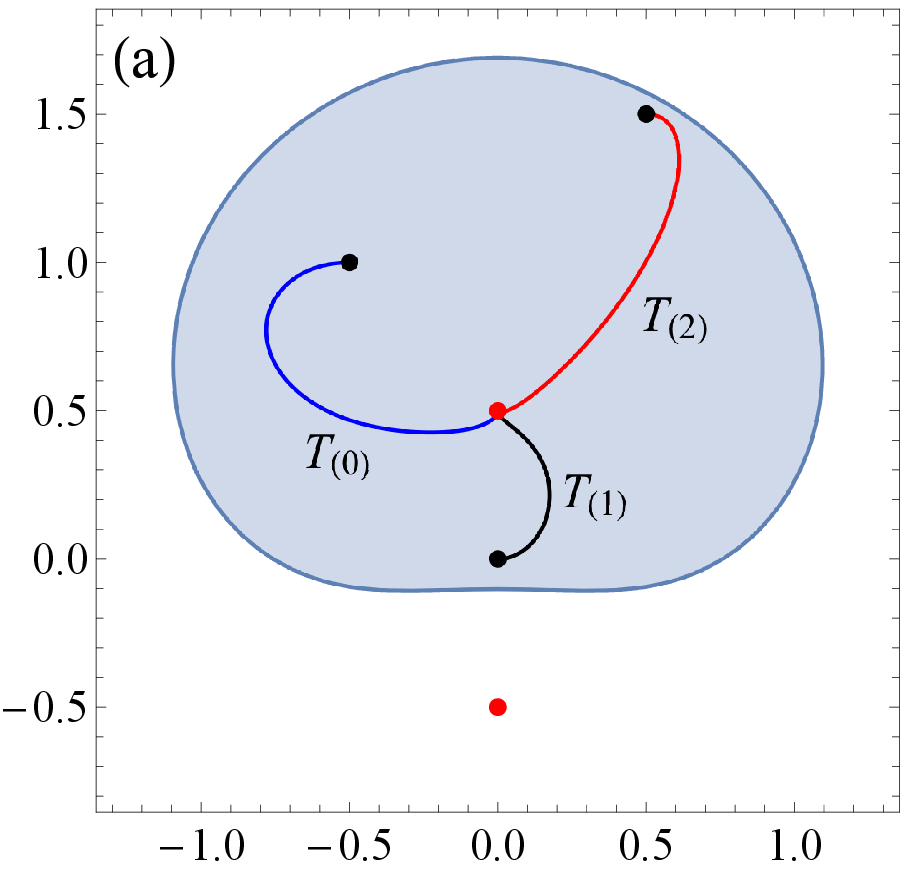}\quad
\includegraphics[width=4cm,height=4cm]{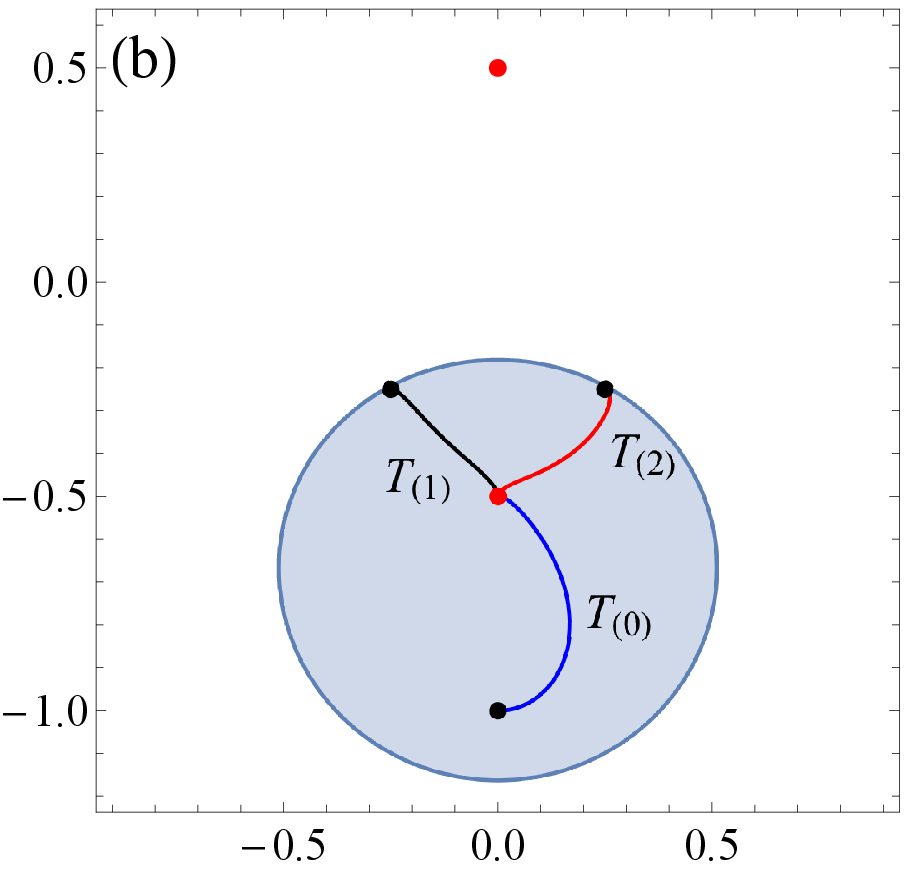}
\caption{\label{fig:examples} (a) Examples of trajectories for particle with charge-mass ratio $\mu=-5$; the other parameters are specified for each trajectory. $T_{(0)}$: $E=-0.1$, $\rho(0)=-1/2$, $z(0)=1$, $\dot{\rho}(0)=-0.73758$ and $\dot{z}(0)=0$. $T_{(1)}$: $E=-0.00001$, $\rho(0)=0$, $z(0)=0$, $\dot{\rho}(0)=1.32396$ and $\dot{z}(0)=0$. $T_{(2)}$: $E=-0.0001$, $\rho(0)=1/2$, $z(0)=3/2$, $\dot{\rho}(0)=0.250857$ and $\dot{z}(0)=0$. (b) Examples of trajectories for particle with charge-mass ratio $\mu=20$; the other parameters are specified for each trajectory. $T_{(0)}$: $E=-0.002$, $\rho(0)=0$, $z(0)=-1$, $\dot{\rho}(0)=0.998144$ and $\dot{z}(0)=0$. $T_{(1)}$: $E=-0.002$, $\rho(0)=-1/4$, $z(0)=-1/4$, $\dot{\rho}(0)=0.423686$ and $\dot{z}(0)=0$. $T_{(2)}$: $E=-0.002$, $\rho(0)=1/4$, $z(0)=-1/4$, $\dot{\rho}(0)=0.423686$ and $\dot{z}(0)=0$. The parameters are $M_{1}=0.4$, $M_{2}=0.2$, and $R=1$.}
\end{figure}

We shall describe two classes of orbits around the BH binary described by BM. The first class of trajectories are orbits with angular momentum $L=0$; this condition implies that the trajectories are confined to a meridional plane, i.e., a plane with $\phi$ constant; for simplicity we take $\phi=0$. Fixing the parameters $M_{1}$, $M_{2}$, $\mu$ and $R$, and given a set of initial conditions $\rho(0)$, $z(0)$ and $\dot{z}(0)$ we can calculate $\dot{\rho}(\lambda)$ and solve the motion equations (\ref{rhozeta}). In Fig.\ \ref{fig:examples} we show some examples of trajectories of particles with negative energy that are confined to the generalized ergosphere. Fig.\ \ref{fig:examples}.a exhibits three trajectories for $\mu=-5$ and Fig.\ \ref{fig:examples}.b exhibits three trajectories for $\mu=20$. Note in these trajectories one of them ends at one of the BHs.

\begin{figure}[ht]
\includegraphics[width=4.2cm,height=4cm]{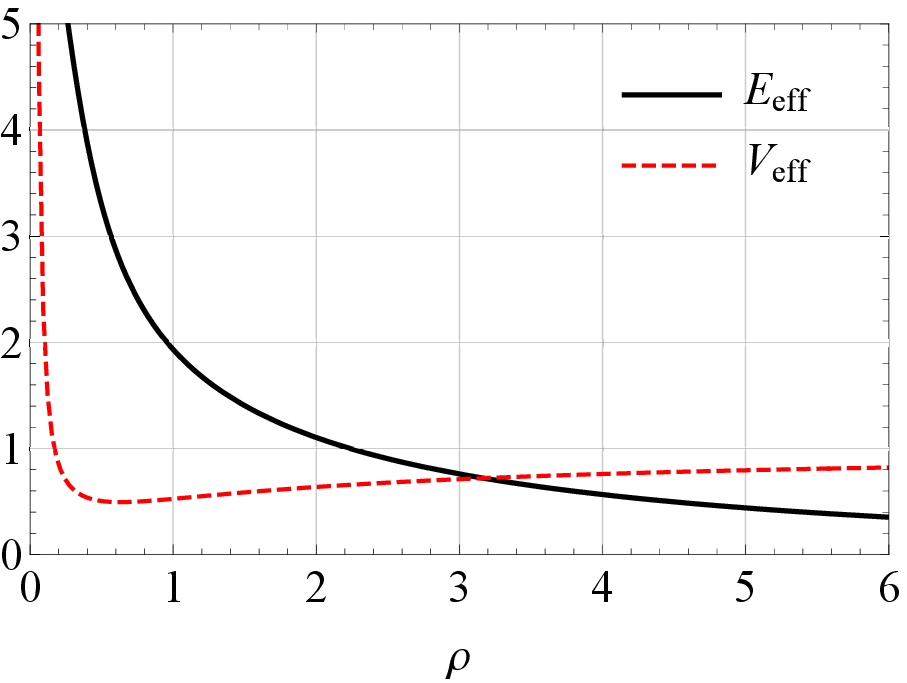}\quad
\includegraphics[width=4cm,height=4cm]{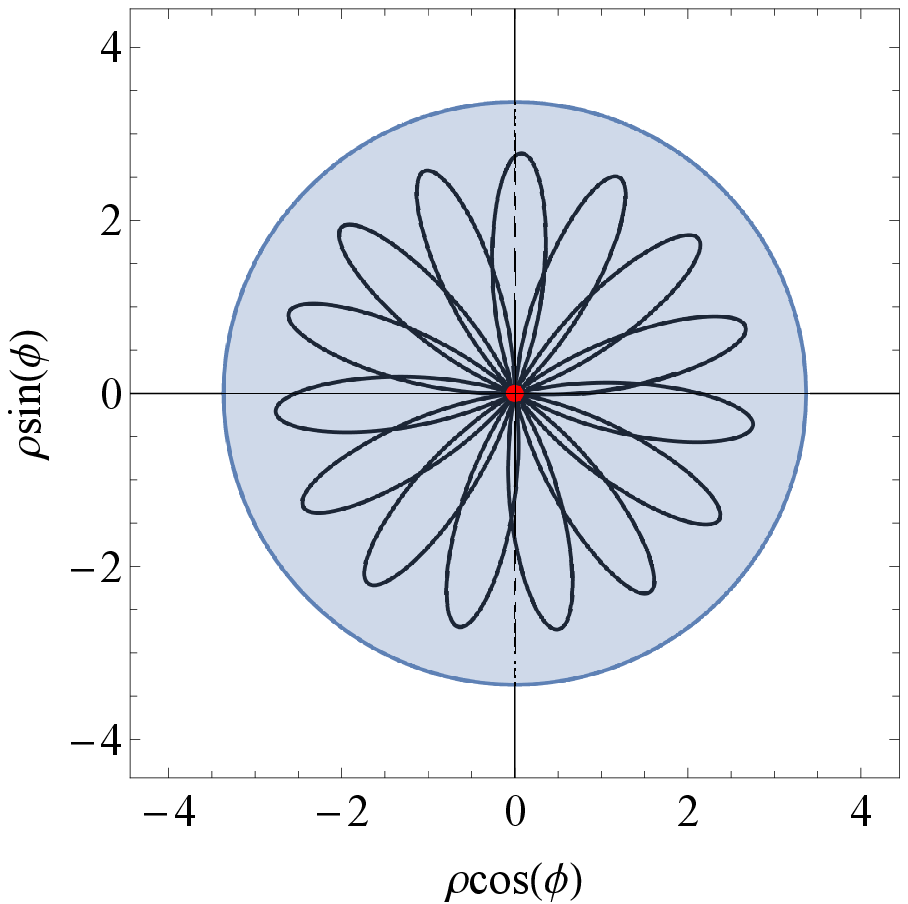}
\caption{Left: Effective energy (black curve) and effective potential (red dashed curve) for $L=0.71$, $E=-0.15$, $M_{1}=0.5$, $M_{2}=0.2$ and $R=1$. Right: example of a trajectory of negative energy at the $z=0$ plane with initial conditions $\rho(0)=2.19858$, $\phi(0)=0$ and $\dot{\rho}(0)=1.22395$. The blue circle corresponds to the generalized ergosphere.}\label{fig:exequatorially}
\end{figure}

\noi The second class of orbits is a projection of geodesics in the plane $z=0$; in general, a particle that initially is located at $z=0$ will not remain in this plane. Since the generalized ergosphere for positive $\mu$ is always contained in the circle depicted by Eq. (\ref{circle}) (see Figs.\ \ref{fig:ergosphere}.f-\ref{fig:ergosphere}.h), then the ergosphere does not reach the plane $z=0$; then we only show the movement for negative $\mu$. Setting $z=0$ in motion Eqs. (\ref{rhozeta}) and energy condition Eq. (\ref{eqrhoz}), the motion in the $z=0$ plane is constrained to the region $E^{2}_{\rm{eff}}(\rho,0) \geq V_{\rm{eff}}(\rho,0)$ where the equality is satisfied for circular orbits. Fig.\ \ref{fig:exequatorially} shows the energy and effective potential for BM at the $z=0$ plane. The corresponding trajectory is also shown in Fig.\ \ref{fig:exequatorially}.

\vspace{-0.2cm}
\section{Penrose process} \label{sec:penrose}
\vspace{-0.2cm}
Now we investigate the possibility of energy extraction from the binary BM. Considering the Penrose process developed in \cite{PenFloyd1971}, extended for RN BHs \cite{DenardoRuffini1972} and lately applied to MP binary BH \cite{Richartz2021}, we addressed the Penrose process for the BM. It consists in sending a charged particle towards the binary BH; at some point, once inside the generalized ergosphere, the particle breaks up into two fragments, one of them escapes to infinity with more energy than the initial one, while the other remains inside the ergosphere until it falls into one of the BHs. We denote the initial particle with subscript $0$, the particle that falls into the BH with subscript $1$ and the particle that escapes with subscript $2$. We consider that the incident particle follows the trajectory $T_{(0)}$, which starts outside the ergosphere and ends inside it, at the break-up point $(\rho_{*}, \phi_{*}, z_{*})$. From the break-up point emerge two particles with trajectories labeled $T_{(1)}$ for the particle with negative energy ($E_{(1)}<0$) that remains inside the ergosphere; and  $T_{(2)}$ that corresponds to the particle escaping to infinity. The $T_{(i)}$ trajectories are timelike paths $x^{\mu}_{i}(\lambda)$ parametrized by the proper time $\lambda$. We denote with $m_{(i)}$, $\mu_{(i)}$, $E_{(i)}$, $L_{(i)}$, and $P^{\mu}_{(i)}$ to the mass, charge-mass ratio, energy per unit mass,  angular momentum per unit mass (with respect to the $z$ axis) and  four momentum {of the $i$-particle}, respectively. The quantities that characterize each particle should fulfill charge, energy and momentum conservation equations. From the charge conservation we have,
 \begin{equation}\label{ccharge}
 \mu_{(0)} m _{(0)}=\mu_{(1)} m _{(1)}+\mu_{(2)} m_{(2)}.
 \end{equation}

On the other hand, if we consider that at the break-up point the four momentum is conserved,
\begin{equation}\label{4momentum}
P^{\nu}_{(0)}=P^{\nu}_{(1)}+P^{\nu}_{(2)}.
\end{equation}

\noi From the temporal component in Eq. (\ref{4momentum}) we have the conservation of the total energy, i.e.
\begin{equation}\label{cenergy}
E_{(0)} m_{(0)}= E_{(1)} m_{(1)}+ E_{(2)} m_{(2)},
\end{equation}

\noi while the spatial components of Eq. (\ref{4momentum}) are the conservations of linear momenta in each component, i.e.,
\begin{equation} \label{linearmomenta}
\begin{aligned}
& m_{(0)}\dot{\rho}_{(0)}=m_{(1)}\dot{\rho}_{(1)}+m_{(2)}\dot{\rho}_{(2)},\\
& m_{(0)}\dot{z}_{(0)}=m_{(1)}\dot{z}_{(1)}+m_{(2)}\dot{z}_{(2)},
\end{aligned}
\end{equation}

\noi where the derivatives $\dot{\rho}_{(i)}$ and $\dot{z}_{(i)}$ should be evaluated at the break-up point. Besides, the conservation of angular momentum is
\begin{equation}
m_{(0)}L_{(0)}=m_{(1)}L_{(1)}+m_{(2)}L_{(2)}.
\end{equation}

Finally there is an additional restriction on the masses $m_{(i)}$; squaring the four momentum (\ref{4momentum}) and using the condition $P^{\mu}_{(1)}P_{\mu(2)}$ (future-pointing timelike vectors)\cite{Richartz2021, Bhat1985}, we have,
\begin{equation}\label{massconstriction}
m^2_{(1)}+m^2_{(2)}<m^2_{(0)}.
\end{equation}

\vspace{-0.2cm}
\section{ENERGY EXTRACTION EFFICIENCY}\label{sec:eficiencia}
\vspace{-0.2cm}
The efficiency $\eta$ of the Penrose process can be defined as the ratio between the output energy (energy of the outgoing particle) and the input energy (energy of the incident particle). From (\ref{cenergy}), we have
\begin{equation}\label{efficiency}
\eta = \frac{E_{(2)}m_{(2)}-E_{(0)}m_{(0)}}{E_{(0)}m_{(0)}} = -\frac{E_{(1)}m_{(1)}}{E_{(0)}m_{(0)}}.
\end{equation}

In order to maximize the efficiency of the process we need to make $E_{(1)}$ as large as possible and $E_{(0)}$ as small as possible. On the other hand the mass of the negative energy fragment $m_{(1)}$ should be as massive as possible in comparison with $m_{(0)}$. In order to deduce the values of the parameters that maximizes the efficiency we choose particular values for the BM parameters $M_{1}$, $M_{2}$, $R$, the break-up point coordinates $(\rho_{*},z_{*})$, and the charge-mass ratio $\mu_{(1)}$. With these considerations we analyze how much energy can be extracted from the BM binary BH.

\vspace{-0.2cm}
\subsection{Maximum efficiency}\label{maxeficiency}
\vspace{-0.2cm}
The minimum energy of the incident particle, that comes from infinity, according to Eq. (\ref{energy}) is $E_{(0)}=1$ and it corresponds to the particle having zero kinetic energy at infinity. On the other hand, with the purpose of maximizing the efficiency, the absolute value of the energy $E_{(1)}$ should be as large as possible, this occurs when the particle $m_{(1)}$ is initially at rest. Recalling that $E_{(1)}<0$, at the break-up point we set
\be \dot{\rho}_{(1)}=\dot{z}_{(1)} =\dot{\phi}_{(1)}=0,\label{rest}\ee

\noi hence, the angular momentum per unit mass and energy per unit mass are, respectively, $L_{(1)}=0$ and
\begin{equation}\label{emin2}
 E_{(1)}(\rho_{*}, z_{*}) =E^{min}_{(1)}(\rho_{*}, z_{*}) = - \mu_{(1)} A_{t*} + \sqrt{f_{*}},
\end{equation}

\noi where $A_{t*}$ and  $f_{*}$ are evaluated at the break-up point $(\rho_{*}, z_{*})$. Now, we determine the restrictions {over} the masses $m_{(0)}$, $m_{(1)}$ and $m_{(2)}$. From the linear momentum conservation Eqs. (\ref{linearmomenta}), we have
\begin{equation}\label{cmomentum2}
\begin{aligned}
m^2_{(2)}=& m^2_{(0)}\frac{\dot{\rho}^2_{(0)}+\dot{z}^2_{(0)}}{\dot{\rho}^2_{(2)}+\dot{z}^2_{(2)}}+m^2_{(1)}\frac{\dot{\rho}^2_{(1)}
+\dot{z}^2_{(1)}}{\dot{\rho}^2_{(2)}+\dot{z}^2_{(2)}}\\
&-2  m_{(0)} m_{(1)} \frac{\dot{\rho}_{(0)}\dot{\rho}_{(1)}+\dot{z}_{(0)}\dot{z}_{(1)}}{\dot{\rho}^2_{(2)}+\dot{z}^2_{(2)}}.
\end{aligned}
\end{equation}

If we consider the condition given by Eq.\ (\ref{rest}), and using  $\left(\dot{\rho}^2_{(i)}+\dot{z}^2_{(i)}\right)$ from Eq.\  (\ref{eqrhoz}), substituting in Eq.\ (\ref{cmomentum2}), yields
\begin{equation}\label{mass2}
m_{(2)}= \sqrt{m^2_{(0)} - 2m_{(0)}m_{(1)} \alpha_{(0)}+m^{2}_{(1)}},
\end{equation}
\noi where

\begin{equation}
\alpha_{(0)}=\frac{1+A_{t*} \mu_{(0)}}{\sqrt{f_{*}}},
\end{equation}

\noi where $f_{*}$ and $A_{t*}$ are evaluated at the break-up point. $\alpha_{(0)}$ can be written as $\alpha_{(0)}=E_{\rm{eff(0)}}/\sqrt{V_{\rm{eff(0)}}}$ using Eqs. (\ref{effveff}) for an initially at rest particle at infinity ($E_{(0)}=1$, $L=0$), hence $\alpha_{(0)}$ is always positive. The MP case is recovered  with $f_{*} \rightarrow 1/U^{2}_{*}$ and  $A_{t*}\rightarrow 1/U_{*} - 1$ where $U_{*}$ is the interaction potential for MP binary BH \cite{Richartz2021}.

From the inequality (\ref{massconstriction}), squaring (\ref{mass2}), and using the fact that the masses are positive, we obtain
\begin{equation}\label{massinequality}
0< m_{(1)}< m_{(0)}\left(\alpha_{(0)}-\sqrt{\alpha^2_{(0)}-1}\right).
\end{equation}

Where a necessary condition for the masses to have real values is $\alpha^2_{(0)}\geq 1$, namely,
\begin{equation}\label{alfa}
\alpha_{(0)}=\frac{1+A_{t*} \mu_{(0)}}{\sqrt{f_{*}}}\geq 1,
\end{equation}

\noi then the values of $\mu_{(0)}$ depend on the location of the break-up point, since $A_{t*}$ is positive (negative), depending if it is evaluated inside (outside) of the region defined by Eq. (\ref{circle}). For the case where the break-up point occurs inside Eq. (\ref{circle}), $\mu_{0}$ is constrained by
\begin{equation}\label{mu1}
\mu_{(0)}\geq \frac{\sqrt{f_{*}}-1}{A_{t*}}.
\end{equation}

Have in mind that the extraction process occurs in this region if $\mu_{(0)}$ is positive and the BH is negatively charged. On the other hand, if the break-up point occurs outside  (\ref{circle}), the values which $\mu_{(0)}$ can take are constrained by
\begin{equation}\label{mu2}
\mu_{(0)}\leq \frac{\sqrt{f_{*}}-1}{A_{t*}},
\end{equation}

\noi and the extraction process takes place if $\mu_{(0)}$ is negative. Note that the change of sign in the inequality (\ref{mu2}) is because $A_{t*}<0$. Then, to maximize the range of $m_{(1)}$ the inequality (\ref{alfa}) must be saturated; this occurs when the inequalities (\ref{mu1}) or (\ref{mu2}) are saturated, i.e., when $\mu_{(0)} \rightarrow  (\sqrt{f_{*}}-1)/A_{t*}$ to the left or to the right according to the sign of $\mu_{(0)}$. In this case, one can choose  $m_{(1)} \rightarrow m_{(0)} $ thus maximizing the ratio $m_{(1)}/m_{(0)}$ that appears in (\ref{efficiency}). Then, the efficiency $\eta$ of the Penrose process in the BM is bounded by
\be \eta < \eta^b = -E_{(1)}^{min}(\rho_{*},z_{*}). \label{emin3}\ee

The efficiency upper bound denoted by $\eta^{b}$ is a function of $\mu_{(1)}$, the break-up point coordinates $(\rho_{*},  z_{*})$ and the BM parameters $M_{1}$, $M_{2}$, $R$ according to Eq. (\ref{emin2}). The effect of varying these parameters is analyzed in the next subsection.
\begin{figure*}[ht]
\includegraphics[width=8.0cm,height=6.0cm]{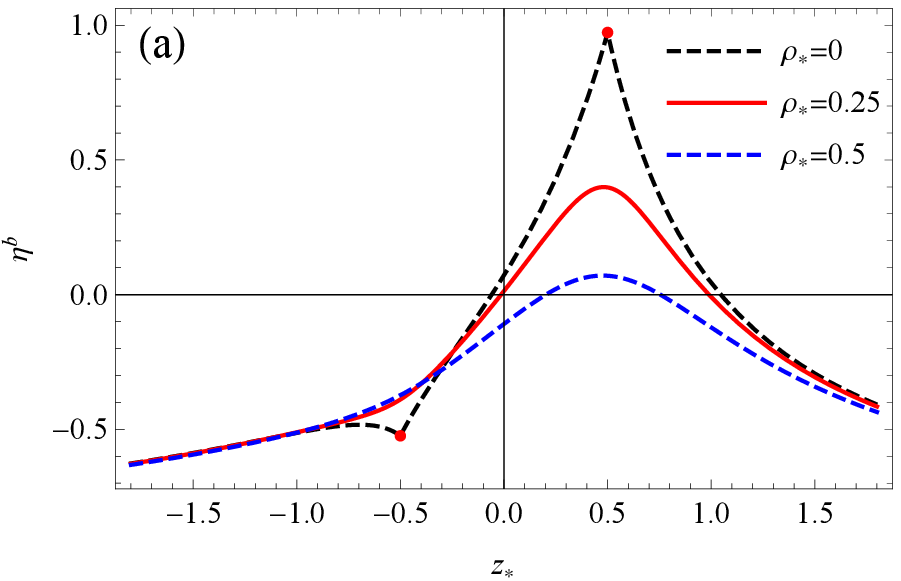}\quad
\includegraphics[width=8.0cm,height=6.0cm]{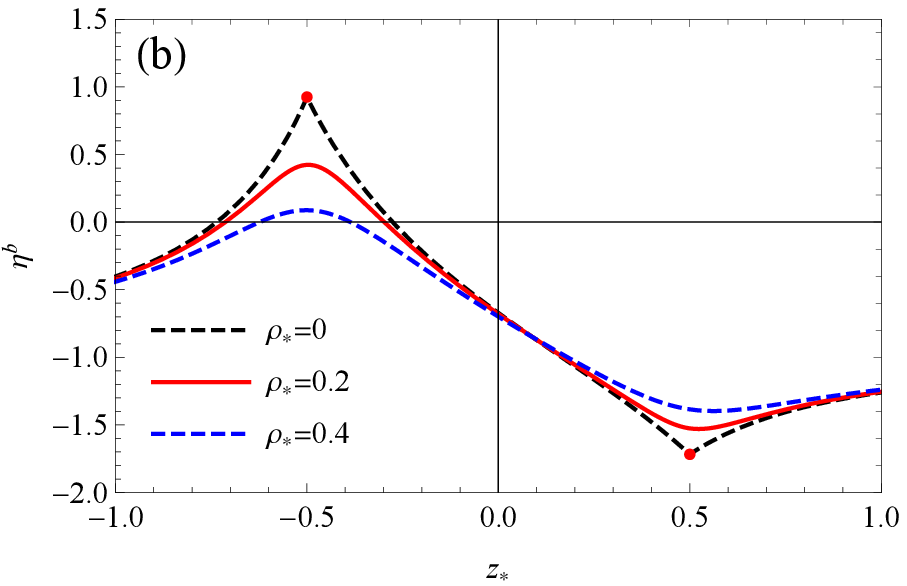}
\caption{\label{fig:eficiencia} It is illustrated the  upper bound efficiency $\eta^{b}$ as a function of $z_{*}$ for fixed $\rho_{*}$ and $M_{1}=0.5$, $M_{2}=0.2$, $R=1$; in (a) $\mu_{(1)}=-1.7$, while in (b) $\mu_{(1)}=3$. In the region where the Penrose process takes place, note that as the break-up point gets closer to the point where the BH is located, the maximum efficiency $\eta^{b}$ increases up to a maximum [see Eqs.\ (\ref{bound1})-(\ref{bound2})]; conversely, if the break-up point moves away  the BH then $\eta^{b} \rightarrow 0$. Negative $\eta^{b}$ occurs if the break-up points are outside the generalized ergosphere.}
\end{figure*}
\begin{figure*}[ht]
\includegraphics[width=7.3cm,height=5.5cm]{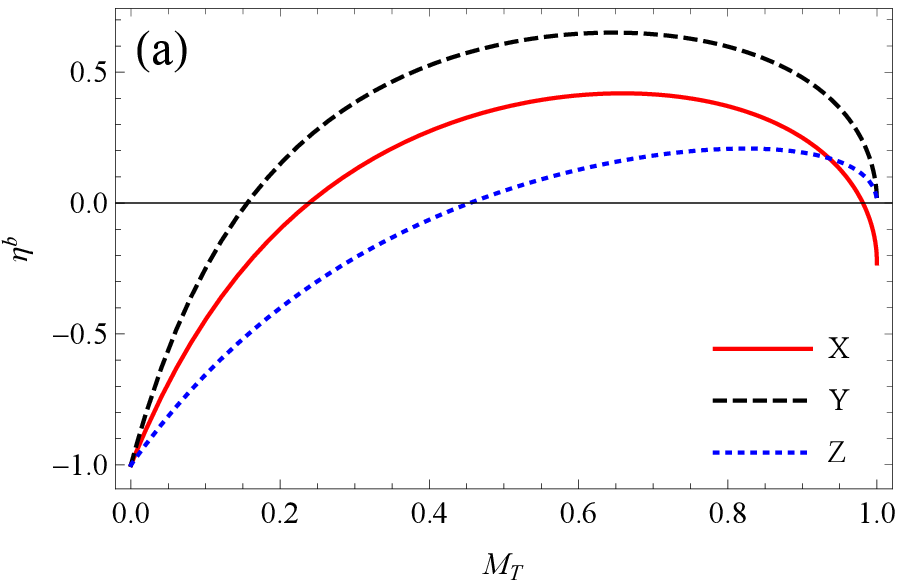}\quad
\includegraphics[width=5.5cm,height=5.5cm]{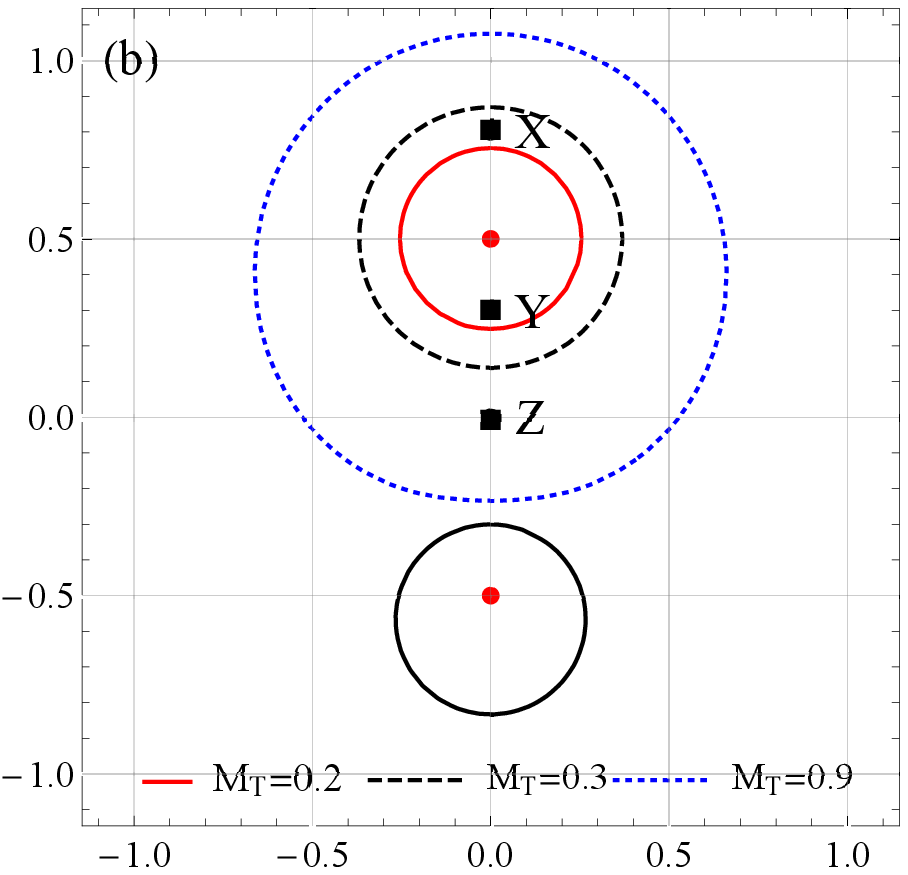}
\caption{It is shown in (a) the  upper bound efficiency $\eta^{b}$ as a function of $M_{T}=M_{1}+M_{2}$ for $M_{R}=M_{2}/M_{1}=0.25$, $R=1$ and $\mu=-1.7$ for selected break-up points, denoted by $X$, $Y$ and $Z$. The values of $M_{Tcrit}$ are $M_{Tcrit}=0.658999$, $M_{Tcrit}=0.645854$ and $M_{Tcrit}=0.823573$ for the break-up points $X$, $Y$ and $Z$, respectively. (b) Generalized ergospheres for selected values of $M_{T}$. The black square markers are the locations of the break-up points $(\rho_{*},z_{*})=(0, 0.8)$, $(\rho_{*},z_{*})=(0, 0.3)$ and $(\rho_{*},z_{*})=(0, 0)$ that are labeled as $X$, $Y$ and $Z$, respectively. Each curve in (a) corresponds to one of the break-up points in (b). The dots represent the BHs.}\label{fig:mtdependence1}
\end{figure*}
\vspace{-0.2cm}
\subsection{Dependence of the maximum efficiency on the parameters.}
\vspace{-0.2cm}
From Eqs.\ (\ref{emin2}) and (\ref{emin3}) the explicit expression of the efficiency upper bound $\eta^b$ is given by
\begin{equation}\label{efficiencybound}
\eta^{b}= \mu_{(1)} A_{t*}-\sqrt{f_{*}}.
\end{equation}

The efficiency bound  $\eta^{b}$ has a linear dependence respect to the charge-mass ratio $ \mu_{(1)}$. The dependence with respect to the break-up point coordinates ($\rho_{*}$, $z_{*}$) can be understood with the help of the energy levels shown in Fig. \ref{fig:level}; the efficiency upper bound increases as the break-up point approaches one of the BHs, i.e., the Penrose process is more efficient if the break-up point is located near one of the horizons. If this is the case, using Eq. (\ref{chargesB}) the upper bound efficiency for $\mu_{(1)}>0$ is
\begin{equation}\label{bound1}
\eta^{b}=\mu_{(1)}\sqrt{\frac{(R-M_{1})^2-M^2_{2}}{(R+M_{1})^2-M^2_{2}}} =- \mu_{(1)}\frac{M_{2}}{Q_{2}},
\end{equation}

\noi and for $\mu_{(1)}<0$ the upper bound is
\begin{equation}\label{bound2}
\eta^{b}=-\mu_{(1)} \sqrt{\frac{(R-M_{2})^2-M^{2}_{1}}{(R+M_{2})^2-M_{1}^2}}=-\mu_{(1)} \frac{M_{1}}{Q_{1}}.
\end{equation}

\noi A relevant characteristic of the efficiency bound,  Eqs.\ (\ref{bound1}) and (\ref{bound2}), is that it depends on the charge-mass ratio $\mu_{(1)}$, and it can be chosen arbitrarily large, {even such that} the resulting efficiency is greater than one, $\eta^{b}>1$, {similarly to the case of one single charged BH} \cite{Bhat1985,Parthasarathy1986,Nucamendi2022} interacting with charged particles.

Moreover, for a given $M_{1}$, the ratio $|M_{2}/Q_{2}|$ in Eq.\ (\ref{bound1})
decreases monotonically in the interval  $0<M_{2}< R-M_{1}$; 
however, as $M_{2}$ approaches $(R-M_{1})$ the ratio $|M_{2}/Q_{2}| \to 0$ and consequently
$\eta^{b}\to 0$. i.e. as the value of the total mass approaches the BH separation distance $R$, $M_{1}+M_{2} \to R$, the maximum efficiency decreases up to zero. The same behavior occurs
for a fixed $M_{2}$ and varying $M_{1}$, see Eq.\ (\ref{bound2}).

From (\ref{efficiencybound}) we can identify two scenarios according to the sign of the charge-mass ratio $\mu$,  {while the sign of $A_{t}$ is defined by the break-up point location}. 

How $\eta^{b}$ depends on the the break-up location, ($\rho_{*}$, $z_{*}$), is illustrated in Fig. \ref{fig:eficiencia}.a for  several values of $\rho_{*}$ and fixed $M_{1}$, $M_{2}$, $R$ and  $\mu_{(1)}<0$. On the other hand, the dependence of the break-up coordinates of $\mu_{(1)}>0$ is illustrated in Fig.\ \ref{fig:eficiencia}.b. The efficiency is negative when the particle that escapes to infinity carries less energy than the incident particle. Note that the extraction process occurs only when the test particle and the BH are of opposite charges.

Since the upper bound efficiency given by Eqs.\ (\ref{bound1}) and (\ref{bound2}) can be written in terms of the BH charges that are such that $|Q_{i}|>M_{i}$, then the upper bound efficiency is less than  $(-\mu_{(1)})$ for arbitrary $M_{1}$, $M_{2}$, and $R$; this is in contrast with the MP case, analyzed in \cite{Richartz2021}, where the upper bound is $(-\mu_{(1)})$. In the case that one of the BH masses is zero, the extraction process occurs if $\mu_{(1)}$ and the BH are oppositely charged and in that case the maximum efficiency is $\eta^{b}=\pm \mu_{(1)}$, for a negative or positively charged BH,  respectively.

To determine how the efficiency depends on the BH masses, we analyze (\ref{efficiencybound}) in terms of the mass ratio $M_{R}=M_{2}/M_{1}$ with $M_{2}<M_{1} $,  total mass $M_{T}=M_{1}+M_{2}$ and fixed $\mu_{(1)}$ and $R=1$, such that $0<M_{R}<1$ and $0<M_{T}<1$. The dependence on the total mass $M_{T}$ for a fixed mass ratio $M_{R}$, $\mu_{(1)}$ with different break-up points is shown in Fig.\ \ref{fig:mtdependence1}. In Fig. \ref{fig:mtdependence1}.a initially the efficiency is negative because the break-up points are located outside the ergosphere and the total mass is small (therefore the ergosphere is small as well); for a fixed  $M_{R}$ the generalized ergosphere gets bigger as the total mass $M_{T}$ increases until a critical value $M_{Tcrit}$, where the generalized ergosphere has its largest size and the upper bound efficiency is maximum. 

When $M_{Tcrit} < M_{T} < R$, the efficiency $\eta^{b}$
behaves similarly than in Eqs.\ (\ref{bound1})-(\ref{bound2});
i.e. it decreases monotonically, even reaching negative values; 
the reason is that as $M_{T}\to R$ the ergosphere shrinks.

The critical value of the total mass $M_{Tcrit}$ is given by 
\begin{equation}
\frac{d}{dM_{T}}\left(\eta^b\right)=0, \quad \frac{d^2}{dM_{T}^2}\left(\eta^b\right)<0, \label{mcritic}
\end{equation}

\noi where $\eta^{b}$ is the upper bound efficiency in Eq. (\ref{efficiencybound}), where $A_{t*}$ and $f_{*}$ have been rewritten in terms of the total mass $M_{T}$ and mass ratio $M_{R}$ using $M_{1}= M_{T}/(M_{R}+1)$ and $M_{2}=M_{R} M_{T}/(M_{R}+1)$.

\begin{figure*}[ht]
\includegraphics[width=7.3cm,height=5.5cm]{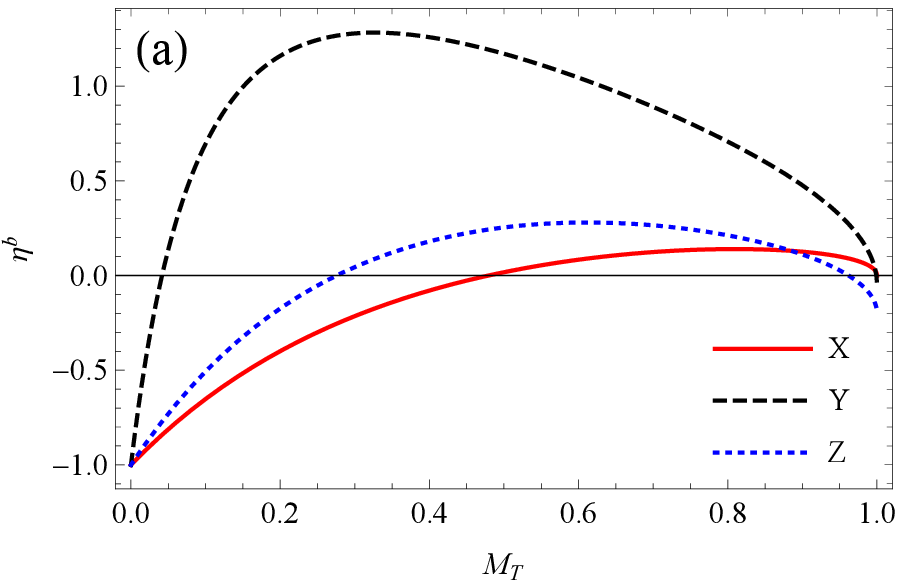}\quad
\includegraphics[width=5.5cm,height=5.5cm]{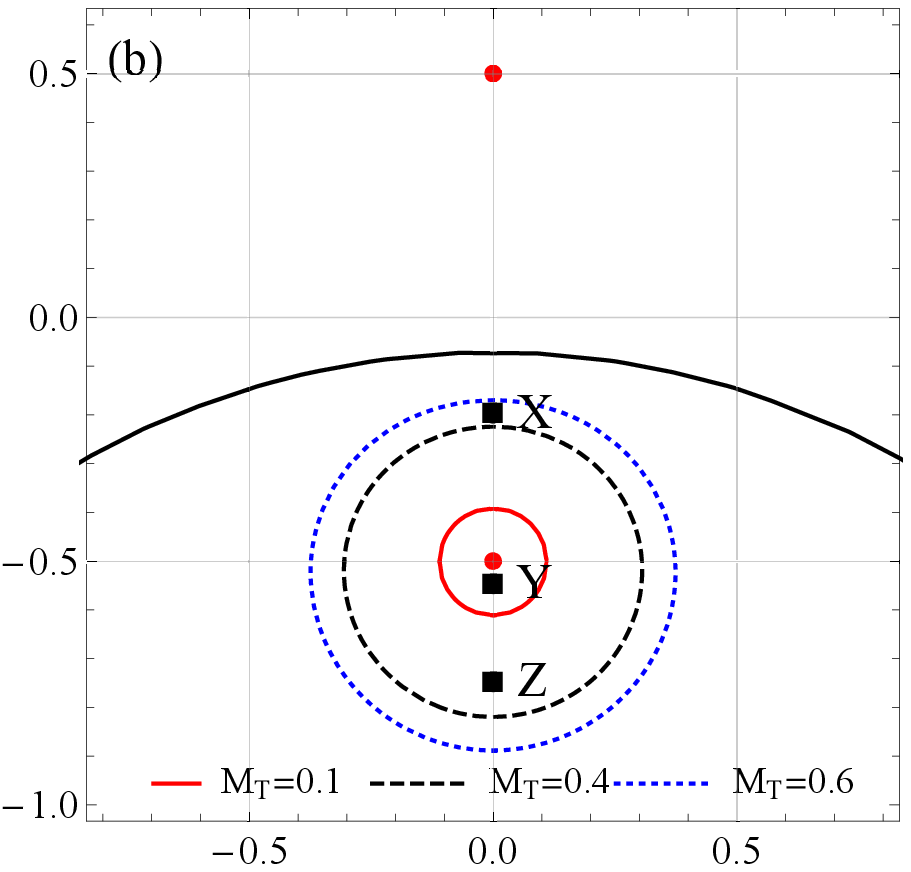}
\caption{It is shown the (a) upper bound  efficiency $\eta^{b}$ as a function of $M_{T}$ for $M_{R}=0.75$, $R=1$ and $\mu=3$ for different break-up points;  $M_{Tcrit}=0.809496$, $M_{Tcrit}=0.326936$ and $M_{Tcrit}=0.611923$ correspond to the break-up points $X$, $Y$ and $Z$, respectively. (b) Generalized ergospheres for different  $M_{T}$. The square markers are the locations of the break-up points $(\rho_{*},z_{*})=(0, -0.2)$, $(\rho_{*},z_{*})=(0, -0.55)$ and $(\rho_{*},z_{*})=(0, -0.75)$ that are labeled $X$, $Y$ and $Z$ respectively. The dots represent the BHs. Each curve in (a) corresponds to one of the break-up points in (b).}\label{fig:mtdependence2}
\end{figure*}

For $\mu_{(1)} > 0$ the ergosphere and upper bound efficiency exhibit the same qualitative behaviour as shown in Fig.\ \ref{fig:mtdependence2}; the critical mass can be calculated with Eqs. (\ref{mcritic}).

\begin{figure*}[ht]
\includegraphics[width=7.3cm,height=5.5cm]{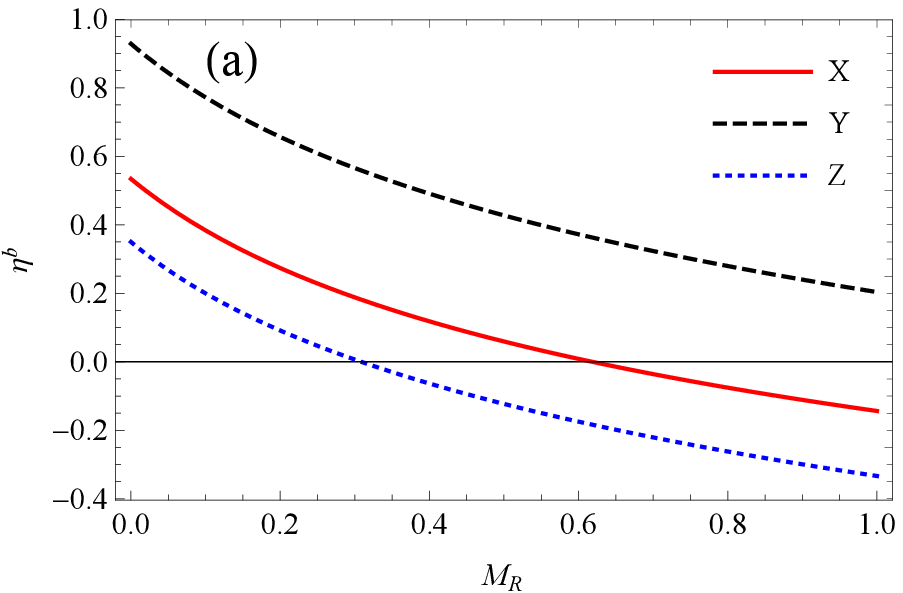}\quad
\includegraphics[width=5.5cm,height=5.5cm]{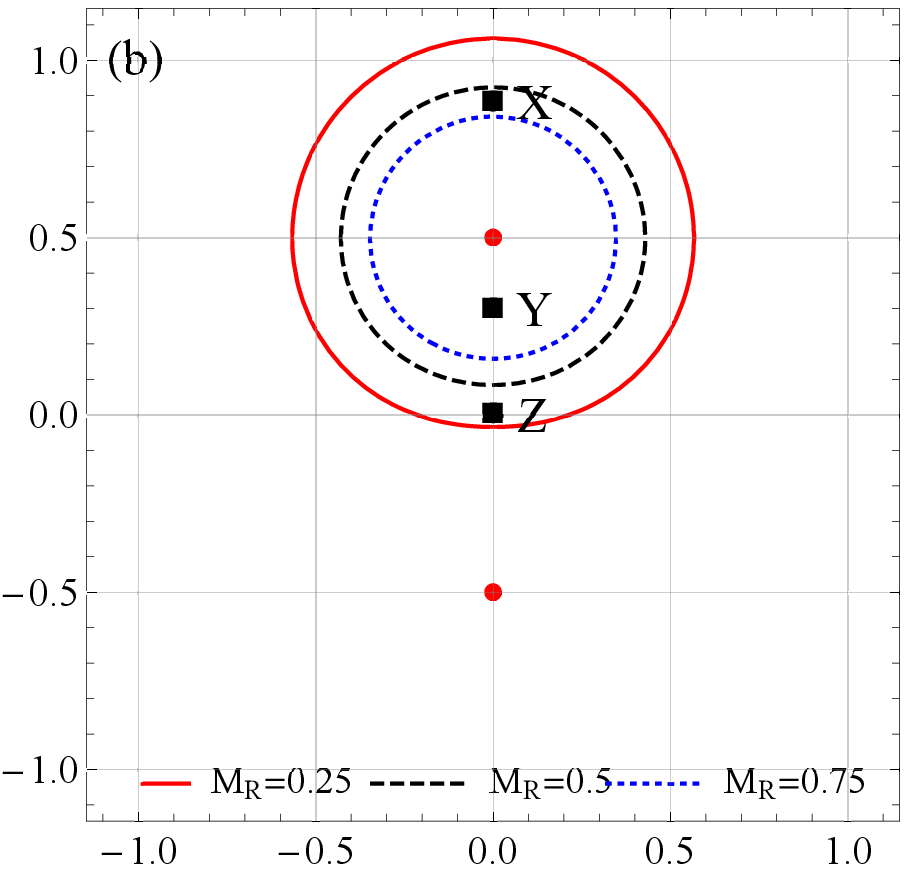}
\caption{It is plotted the (a) upper bound  efficiency $\eta^{b}$ as a function of $M_{R}$ for $M_{T}=0.5$, $R=1$ and $\mu=-1.7$ for selected break-up points. (b) Generalized ergospheres for different $M_{R}$. The square markers  are the locations of the break-up points $(\rho_{*},z_{*})=(0, 0.88)$, $(\rho_{*},z_{*})=(0, 0.3)$ and $(\rho_{*},z_{*})=(0, 0)$labeled as $X$, $Y$ and $Z$, respectively. The dots represent the BHs. Each curve in (a) corresponds to one of the break-up points in (b).}\label{fig:mtdependence3}
\end{figure*}

\begin{figure*}[ht]
\includegraphics[width=7.3cm,height=5.5cm]{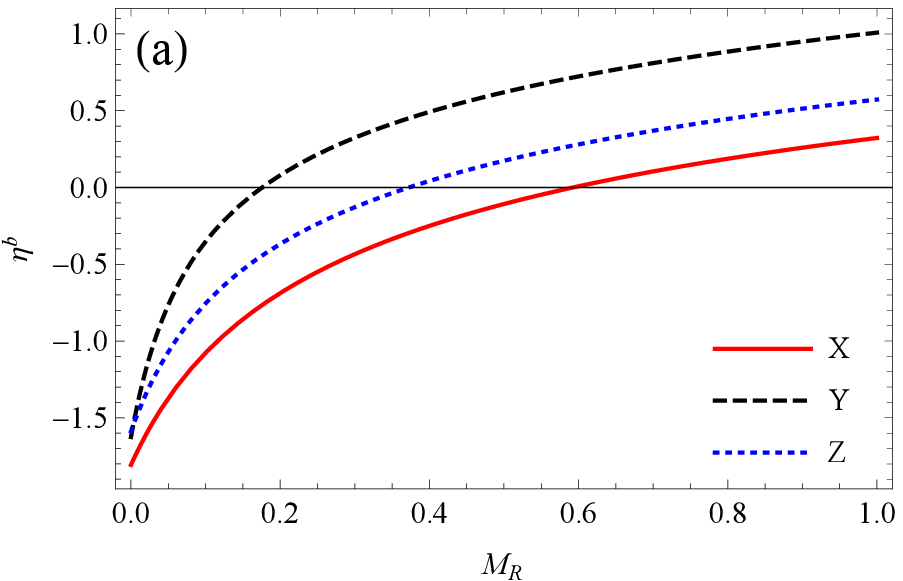}\quad
\includegraphics[width=5.5cm,height=5.5cm]{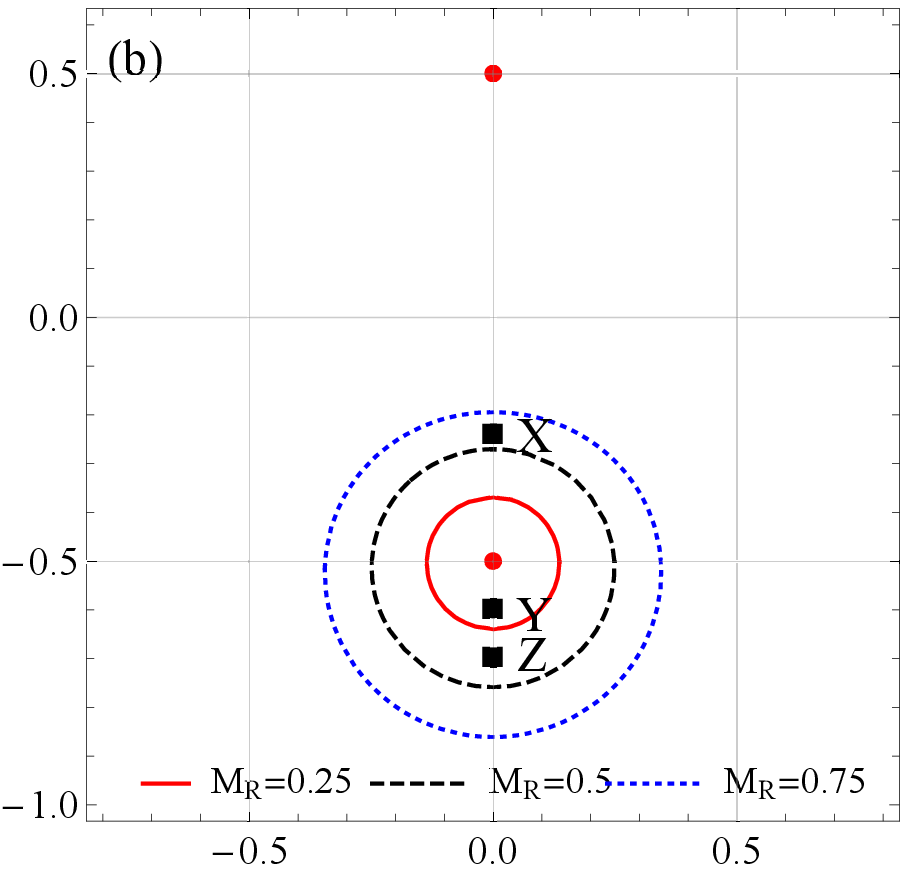}
\caption{It is illustrated in (a) the upper bound  efficiency $\eta^{b}$ as a function of $M_{R}$ for $M_{T}=0.5$, $R=1$ and $\mu=3$ for selected break-up points. (b) Generalized ergospheres for different $M_{R}$. The square markers  are the locations of the break-up points $(\rho_{*},z_{*})=(0, -0.24)$, $(\rho_{*},z_{*})=(0, -0.6)$ and $(\rho_{*},z_{*})=(0, -0.7)$ with labels $X$, $Y$ and $Z$ respectively. The dots represent the BHs. Each curve in (a) corresponds to one of the break-up points in (b).}\label{fig:mtdependence4}
\end{figure*}

The mass ratio $M_{R}$ for $M_{T}=1/2$ and $\mu_{(1)} < 0 $, for different break-up points is shown in Fig. \ref{fig:mtdependence3}. In this case the ergosphere and upper bound $\eta^{b}$ decrease as $M_{R}$ increases, i.e., the ergosphere and upper bound efficiency decrease when the masses $M_{1}$  and $M_{2}$ tend to the same value. The maximum of the upper bound efficiency $\eta^{b}$ occurs when $M_{R} \to 0$, i.e., when $M_{2}\to 0$ but less or equal to $-\mu_{(1)}$. In Fig. \ref{fig:mtdependence4} is shown $\eta^b$ as a function of $M_{R}$ for $\mu_{(1)}>0$. In this case the upper bound efficiency $\eta^{b}$ and the ergosphere increase as  $M_{R}$ increases and the maximum $\eta^{b}$ for an arbitrary break-up point occurs when $M_{R}\to 1$.
\vspace{-0.2cm}
\subsection{Examples of the Penrose process}
\vspace{-0.2cm}
Once we have described the efficiency $\eta^{b}$, in this section some concrete examples of the Penrose process in the BM are given.  In these examples the efficiency approaches the theoretical maximum described by  Eq. (\ref{bound1}) for test particles with $\mu_{(1)}>0$ and Eq. (\ref{bound2}) for test particles with $\mu_{(1)}<0$. First we fix $M_{1}$, $M_{2}$, and $R$, and the charge mass ratio $\mu_{(1)}$ for a particular break-up point $(\rho_{*},z_{*}, \phi_{*})$ located inside the generalized ergosphere. According to the analysis in Sec. \ref{maxeficiency} we set $m_{(0)}=1$, $L_{(0)}=L_{(1)}=L_{(2)}=0$, which means that the trajectories are confined to the meridional plane $\phi=\phi_{*}$. The energy $E_{(1)}$ is determined by Eq. (\ref{emin2}) and we set $E_{(0)}=1$. The two scenarios for the process correspond to negative and positive charged test particle $\mu_{(1)} $; the charge mass ratio $\mu_{(0)}$ is bounded by two different limits depending where the break-up point ($\rho_{*}$, $z_{*}$) occurs according to Eqs. (\ref{mu1}) and (\ref{mu2}).

If the break-up point is located outside the region bounded by (\ref{circle}), then according to Eqs. (\ref{massinequality}) and (\ref{mu2}), we choose
\begin{equation}\label{epsilon1}
\mu_{(0)} = \frac{\sqrt{f_{*}}-1}{A_{t*}}-\epsilon_{1},
\end{equation}
\begin{equation}\label{nu1}
m_{(1)}=1-\nu_{1},
\end{equation}
\begin{figure*}[ht]
\includegraphics[width=6cm,height=6cm]{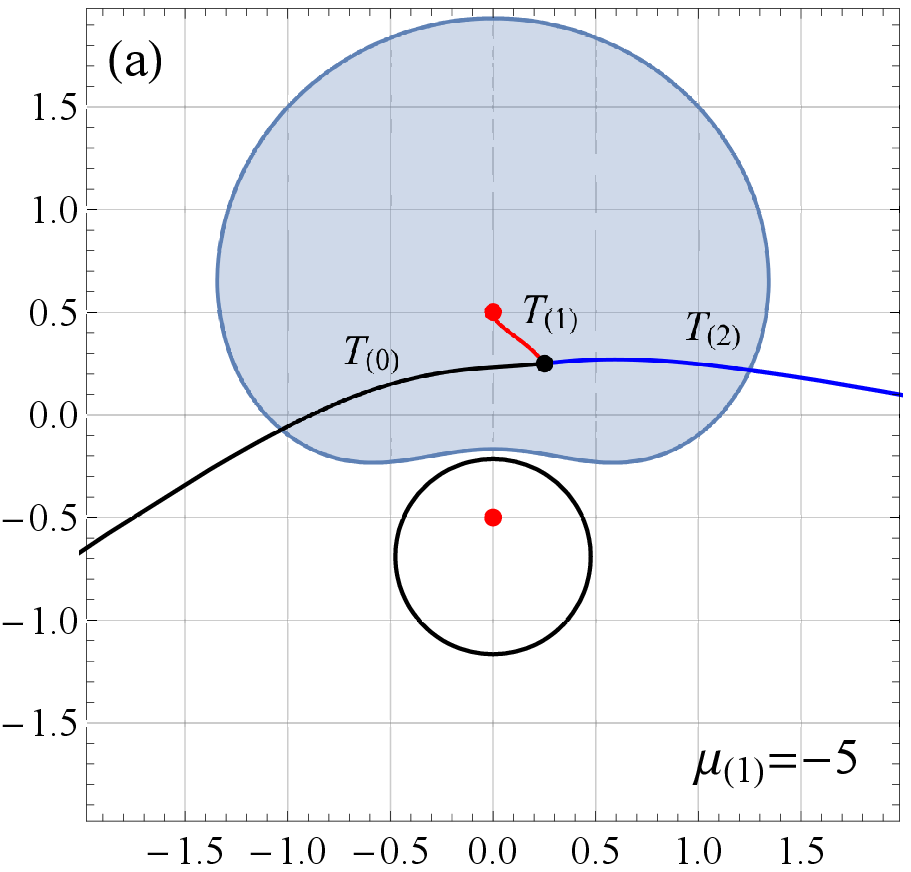}\quad
\includegraphics[width=6.0cm,height=6.0cm]{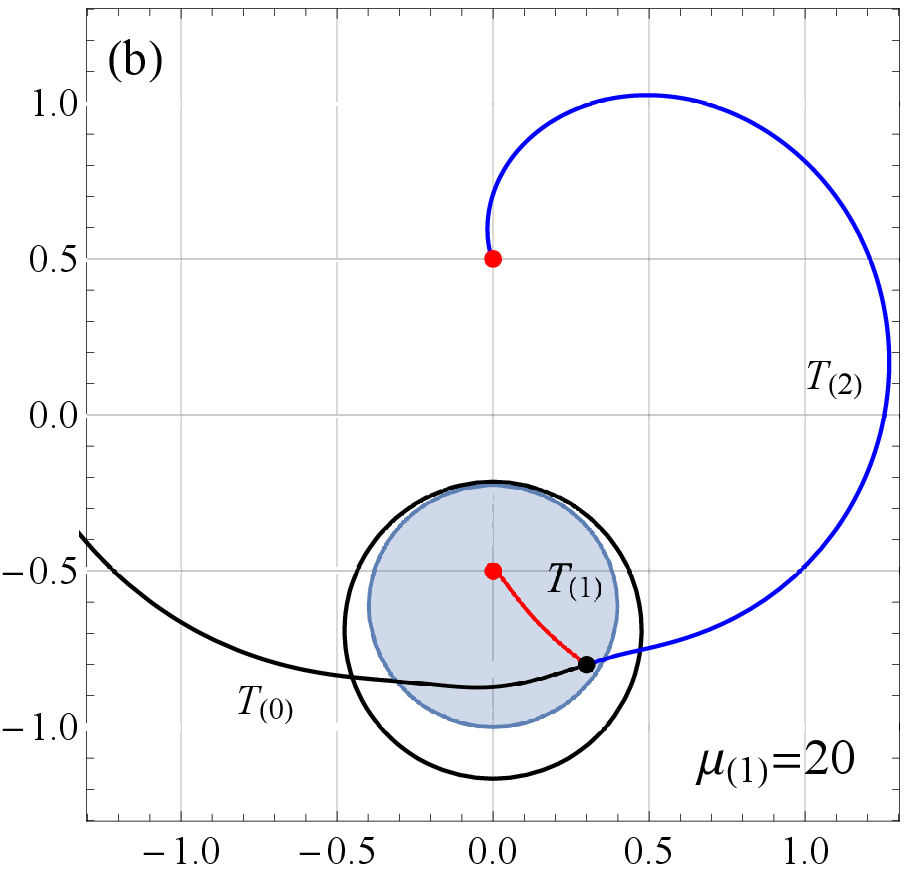}
\caption{Examples of Penrose process with the maximum efficiency in a BM. The efficiencies of the processes are 90$\%$ of the theoretical maximum. In both panels the incoming trajectory $T_{(0)}$ (black curve) splits (at the black point) into the negative energy trajectory $T_{(1)}$ (red curve) and the trajectory of the particle that escapes $T_{(2)}$. The Bonnor parameters in both cases are fixed as $M_{1}=0.5$, $M_{2}=0.2$, $R=1$, the parameter $\nu$ and $\epsilon$ are fixed as $\nu_{1,2}=10^{-2}$, $\epsilon_{1,2}=10^{-5}$, the charge-mass ratio, the break-up point and the angle $\theta_(0)$ are fixed, respectively, as: (a) $\mu_{(1)}=-5$, $(\rho_{*},z_{*})=(1/4, 1/4)$ and $\theta_{(0)}=0.0872665$; (b) $\mu_{(1)}=20$, $(\rho_{*},z_{*})=(0.3, -0.8)$ and $\theta_{(0)}=0.296706$. The initial values of the parameters that generate these trajectories are shown in Tables \ref{tabla1} and \ref{tabla2}. The respective  upper bound efficiency is $\eta^{b}=1.20854$ (left panel) and $\eta^b = 0.208203$ (right panel).}\label{trajectorymax}
\end{figure*}
\noi where $\epsilon_{1}$ and $\nu_{1}$ are small and positive. Substituting Eqs.\ (\ref{epsilon1}) and (\ref{nu1}) into Eq. (\ref{massinequality}), we find that $\epsilon_1$, and $\nu_1$, satisfy
\begin{equation}\label{nu1inequality}
\nu_{1} > \frac{A_{t*} \epsilon_{1}}{\sqrt{f_{*}}}\left(\sqrt{1-\frac{2\sqrt{f_{*}}}{A_{t*}\epsilon_{1}}}+1\right).
\end{equation}

On the other hand, if the break-up point occurs inside of the circle (\ref{circle}), then according to (\ref{massinequality}) and (\ref{mu1}), we choose
\vspace{-0.1cm}
\begin{equation}\label{epsilon2}
\mu_{(0)} = \frac{\sqrt{f_{*}}-1}{A_{t*}}+\epsilon_{2},
\end{equation}

\vspace{-0.1cm}
\begin{equation}\label{nu2}
m_{(1)}=1-\nu_{2},
\end{equation}

Substituting these expressions into (\ref{massinequality}), we find that $\epsilon_{2}$ and $\nu_{2}$ satisfy
\vspace{-0.1cm}
\begin{equation}\label{nu2inequality}
\nu_{2}> \frac{A_{t*}\epsilon_{2}}{\sqrt{f_{*}}}\left(\sqrt{1+\frac{2\sqrt{f_{*}}}{A_{t*}\epsilon_{2}}}-1\right),
\end{equation}

\noi the charge mass ratio $\mu_{(2)}$ and energy per unit mass $E_{(2)}$ can be determined from Eqs. (\ref{ccharge}) and (\ref{cenergy}).

For a given angle between the velocities, $\theta_{(0)}=\arg(\dot{\rho}_{(0)}+i\dot{z}_{(0)}$),  Eqs. (\ref{rhoz1}) and (\ref{rhoz2}) can be used to determine $\dot{\rho}_{(0)}$ and $\dot{z}_{(0)}$ at the break-up point.
According to (\ref{rest}), we have $\dot{\rho}_{(1)}=\dot{z}_{(1)}=0$ for the negative energy fragment. For the incident particle, if the break-up point is outside the circular region (\ref{circle}) and $E_{(0)}=1$ then
\begin{equation}\label{rhoz1}
\dot{\rho}^{2}_{(0)}+\dot{z}^{2}_{(0)} =\frac{ \big(1+A_{t*}(1-\epsilon_{1})\big)^{2}}{e^{2\gamma_{*}}} - \frac{f_{*}}{e^{2\gamma_{*}}},
\end{equation}

\noi while if the break-up point is inside  (\ref{circle}) we have,
\begin{equation}\label{rhoz2}
\dot{\rho}^{2}_{(0)}+\dot{z}^{2}_{(0)} =\frac{ \big(1+A_{t*}(1+\epsilon_{2})\big)^{2}}{e^{2\gamma_{*}}}  - \frac{f_{*}}{e^{2\gamma_{*}}}.
\end{equation}

While from linear momentum conservation (\ref{linearmomenta}) $\dot{\rho}_{(2)}$ and $\dot{z}_{(2)}$ are calculated.
 
With this setting of parameters the trajectories $T_{(0)}$, $T_{(1)}$ and $T_{(2)}$  can be completely determined and the efficiency of the Penrose process is given by $\eta_{1,2} = (1-\nu_{1,2})\eta^{b}$. In order to maximize the efficiency we choose $\nu_{1,2}$ as small as possible. However, $\epsilon_{1,2}$ and $\nu_{1,2}$ should not be zero because the inequalities (\ref{nu1inequality}) or (\ref{nu2inequality}) cannot be saturated; but it is possible, in principle, to set infinitesimally small values for $\epsilon_{1,2}$ and $\nu_{1,2}$. According to these considerations we explore two specific  examples where the efficiency corresponds to $90\%$ for a break-up point located outside and inside of the region bounded by (\ref{circle}) resulting that the extraction process occurs with the BH with positive or negative charge, respectively.

Fig.\ \ref{trajectorymax}.a shows the Penrose extraction from a BH with negative charge. The BM parameters $M_{1}$, $M_{2}$, and $R$ are fixed. It corresponds to the extraction process where the break-up point occurs outside (\ref{circle}) and the charge-mass ratio of the negative energy fragment (red line) is negative $\mu_{(1)}$  (e. g. Fig.\ \ref{fig:ergosphere}.c), the setting of these parameters enables us to determine the trajectories $T_{(1)}$, having in mind that the energy $E_{(1)}$ associated to the trajectory $T_{(1)}$,  Eq. (\ref{emin1}), is the minimum possible and the trajectories $T_{(0)}$ and $T_{(2)}$ are fully described once we fix $\nu_{1,2}$, $\epsilon_{1,2}$, and $\theta_{(0)}$. On the other hand, in Fig.\ \ref{trajectorymax}.b it is shown the Penrose process where the break-up point occurs inside the region bounded by (\ref{circle}) for a negatively charged particle. 
We can highlight two different features of the Penrose process in the BM that are in contrast with the MP binary: first, the process of extraction can occur for positive or negative charged test particles inside the generalized ergosphere that surrounds the BH with charge of opposite sign to the one of the test particle. So, any observer can recover information of the energy extraction only when the particle that gains energy escapes back to infinity. Second, in some scenarios the particle escaping with more energy can be trapped by the other BH, as shown in Fig.\ \ref{trajectorymax}.b; in this case the observer at infinity will not receive any information of the energy extraction process. As far as we know the second phenomenon has not been reported in the MP binary BH. The parameters that generate these examples satisfy the conservations Eqs. (\ref{ccharge})-(\ref{massconstriction}) and are given in Tables \ref{tabla1} and \ref{tabla2}.
\begin{table}[ht]
\centering
\caption{Initial values that generate the trajectories $T_{(0)}$, $T_{(1)}$ and $T_{(2)}$ in the Penrose process shown in Fig. \ref{trajectorymax}.a. The derivatives $\dot{\rho}_{(i)}$ and $\dot{z}_{(i)}$ are evaluated at the break-up point.} \label{tabla1}
\begin{ruledtabular}
\begin{tabular}{c c c c c c c c}
$i$ & $m_{(i)}$ & $\mu_{(i)}$
 &$E_{(i)}$  &$L_{(i)}$ &$\dot{\rho}_{(i)}$ & $\dot{z}_{(i)}$& \\ \hline
0 & 1                &2.70371& 1 & 0 & 1.45015 & 0.126872\\
1 & 0.9              & -5 & -1.20854 & 0 & 0 & 0 \\
2 & 0.0998852 & 72.1199 & 20.9008& 0 & 14.5182 & 1.27018
\end{tabular}
\end{ruledtabular}
\end{table}
\begin{table}[ht]
\centering
\caption{Initial values that generate the trajectories $T_{(0)}$, $T_{(1)}$ and $T_{(2)}$ in the Penrose process shown in Fig. \ref{trajectorymax}.b. The derivatives $\dot{\rho}_{(i)}$ and $\dot{z}_{(i)}$ are evaluated at the break-up point.} \label{tabla2}
\begin{ruledtabular}
\begin{tabular}{c c c c c c c c}
 $i$ & $m_{(i)}$ & $\mu_{(i)}$
 &$E_{(i)}$  &$L_{(i)}$ &$\dot{\rho}_{(i)}$ & $\dot{z}_{(i)}$& \\ \hline
0 & 1                &2.70371& 1 & 0 & 1.45015 & 0.126872\\
1 & 0.9              & -5 & -1.20854 & 0 & 0 & 0 \\
2 & 0.0998852 & 72.1199 & 20.9008& 0 & 14.5182 & 1.27018
\end{tabular}
\end{ruledtabular}
\end{table}

\vspace{-0.2cm}
\subsection{Stability of the Bonnor metric}
\vspace{-0.2cm}

In \cite{Bernard2019} it is studied the effect of perturbing a black hole  binary (BHB) with a massless scalar field and it is shown that initially the dominant mode is monopolar and once this initial mode dies away there arises an exponentially damped sinusoid mode. It is as well find out that
BHB possesses global quasinormal modes whose ringdown parameters do not depend on the initial conditions but only on the masses and the separation, and that relaxation time scale increases with the separation. Moreover simulations indicate that the perturbations die away in time.  Besides, \cite{Wong2019} agrees in the exponential decay of a massless scalar perturbation, however indicating that for confined BHB there might be mechanisms that could trigger instabilities. More analysis is required for other kind of perturbations to elucidate general conditions under which BHB are stable systems.

Regarding stability of the binary BM against  electromagnetic perturbations and
specifically  how energy extraction affects stability the following considerations point to the stability of the BM: since the generalized ergosphere never encircles
both BH, then the perturbation affects only one of the BH, the one with charge opposite to the test particle; therefore we can apply stability criteria valid for one single charged BH, i.e.
a Reissner-Nordstr\"om BH, that we know is stable;
and then we guess that the BM remains stable regarding energy extraction. 
The situation would be very different if the system be under some kind of confinement
that could stimulate superradiance effects; in any case a deeper analysis is required to elucidate the BM stability.

\section{Conclusions}
\vspace{-0.2cm}
We have analyzed in detail the possibility of energy extraction from the Bonnor BH binary (BM), that describes two oppositely charged BH  kept apart by a strut \cite{Bonnor1979,Cabrera2011} that prevents the two BH collide; this is in contrast to the Majumdar-Papapetrou (MP) binary
where the gravitational and electromagnetic forces are balanced.

We determined a generalized ergosphere that depends on the parameters of the BM and the charge-mass ratio $\mu$ of the test particle and we showed that energy extraction is possible; the sign of the electric potential $A_{t}$ is defined by the point $(\rho,z)$ where it is evaluated. A first difference with the MP case is that the generalized ergosphere exists for positive and negatively charged test particles $\mu$. Another remarkable difference with respect to the MP case is that the ergosphere cannot include both BHs, but only one; the ergosphere encloses the BH with charge opposite to the one of the  test particle. We found that for some  initial conditions  the particle that escapes with more energy is trapped by the other BH and in this case the observer at infinity would not receive information of the energy extraction. 

We studied the conditions that optimize the efficiency of the process.
The efficiency is enhanced if the break-up point is located near the horizon of the BH charged oppositely to the test particle. We determined  the total mass $M_{T}=M_{1}+M_{2}$ and mass ratio $M_{R}=M_{2}/M_{1}$ that renders the highest efficiency $\eta^b$.  The behavior of $\eta^b$ as a function of $M_{R}$ depends on the sign of $\mu$. If $\mu<0$ then $\eta^b$ decreases when $M_{R}\to 1$; while if $\mu>0$ then $\eta^b$ increases when $M_{R}$ approaches $1$. Moreover the maximum efficiency $\eta^b$ does not increase monotonically as $M_{T}$ increases, but there is a certain $M_{Tcrit}$ such that for $M_{T}>M_{Tcrit}$, $\eta^b$ decreases, and even can reach negative values. The upper bound efficiency in the BM is always smaller than the MP one \cite{Richartz2021}.

Due to the vast recent observations reported by the LIGO-Virgo Collaboration, so far it has been able to identify multiple candidates for compact binary systems. We believe that the study of the Penrose process in BM  contributes to the understanding of actual BHBs
as it extends the analysis carried out for a MP BH in \cite{Richartz2021}. 
Further analysis of BHB would give relevant information regarding other astrophysical aspects \cite{Misner1972, Bekenstein1973, Brito2015}, for instance there are proposals that link compact binary systems with superradiance \cite{Rosa2015, Wong2019}. Moreover, the magnetic variant of the Penrose process that takes into account the combined influence of external magnetic field and the rotation of a BH  seems to be connected to the origin of accretion disks where the energy extraction into jets can befall, or even  the generation of ultra-high energy cosmic rays \cite{Kolos2021, Kolos2020}.  In this direction we aim to develop further research of the magnetic Penrose process in BHB.



\vspace{-0.5cm}
\section{Acknowledgments}
\vspace{-0.2cm}
NB acknowledges partial support by CONACyT Project No. 284498. ICM acknowledges financial support of SNI-CONACyT, M\'exico, grant CVU No. 173252. AB acknowledges financial support by CONACyT, M\'exico, through the PhD Scholarship with CVU No. 933515.

\end{document}